# Optimal charging guidance strategies for electric vehicles by considering dynamic charging requests in a time-varying road network

**Yongxing Wang, Jun Bi** *

*School of Traffic and Transportation, Beijing Jiaotong University, Beijing 100044, China*

**A B S T R A C T**

Electric vehicles (EVs) have enjoyed increasing adoption because of the global concerns about the petroleum dependence and greenhouse gas emissions. However, their limited driving range fosters the occurrence of charging requests deriving from EV drivers in urban road networks, which have significant uncertain characteristic from time dimension in the real-world situation. To tackle the challenge brought by the dynamic charging requests, this study is devoted to proposing optimal strategies to provide guidance for EV charging. The time-varying characteristic of road network is further involved in the problem formulation. Based on the charging request information, we propose two charging guidance strategies from different perspectives. One of the strategies considers the travel demands of EV drivers and uses the driving distance as the optimization criterion. In contrast, the other strategy focuses on the impacts of EV number on the charging station operation and service satisfaction. The reachable charging stations with minimum vehicle number are selected as the optimal ones. More importantly, both the strategies have the ability to ensure the reachability of selected charging stations in a time-varying road network. In addition, we conduct simulation examples to investigate the performance of the proposed charging guidance strategies. Besides, the insights and recommendations on application scenarios of the strategies are introduced according to the simulation results under various parameter scenarios.

*Keywords:*

Electric vehicle
Charging guidance strategies
Dynamic charging requests
Time-varying road network

---

* Corresponding author. Address: Beijing Jiaotong University, School of Traffic and Transportation, NO.3 Shangyuancun Street, Beijing 100044, China. Tel.:+86 134 8881 2321 Fax: +86 0105168414
*E-mail addresses*: bilinghc@163.com (J. Bi), yx_wang@bjtu.edu.cn (Y. Wang)

## 1. Introduction

The dependence of human society on petroleum has contributed to serious environmental and energy problems. The transportation sector is one of the major economic industries that contribute to energy consumption and greenhouse gas emissions. According to the investigation conducted by the International Energy Agency, the energy consumption of the transportation sector accounts for 28% of the global energy consumption and is responsible for 23% of the global greenhouse gas emissions (International Energy Agency, 2017). Given the public concern on climate change and advances in battery technologies, electric vehicles (EVs) have been introduced as a promising solution for the problem of dependency on fossil fuels and increasing greenhouse gas emissions (Rezvani et al., 2015). However, unlike conventional internal combustion engine vehicles, EVs have the relatively short driving range due to the limited capacity of batteries. The drivers often need to recharge their vehicles during trips to successfully reach the destinations. Moreover, the charging stations for EVs are considerably less popular than gas stations. These disadvantages increase driver range anxiety, that is, the fear of depleting battery energy en route (Melliger et al., 2018). In order to help drivers to select suitable charging stations and alleviate their range anxiety, a smart charging service would be developed to provide guidance for EV charging. Through such a service, EV drivers send charging requests to the charging operating centre when the battery energy of their vehicles is insufficient to reach the destinations, and the centre provides feedback to the drivers, which is the optimal charging station selection according to the information from the drivers' charging requests (Wang et al., 2018b). To realize the smart charging service, the charging guidance strategies based on charging request information need to be developed. Furthermore, in the real-world travel situation, the traffic condition on a road network often has the time-varying characteristics, which would influence the route and charging station selection for EVs (Gendreau et al., 2015). Thus, besides the charging requests, the time-varying characteristics of road network should be considered in the charging guidance strategies. More importantly, the dynamic characteristic intrinsic to the charging requests has substantial impacts on the strategies, which would further increase the difficulties to deal with the charging requests. Note that, large-scale charging behaviours with dynamic characteristic would exert significant impacts on the operation efficiency of charging stations. Therefore, given the widespread adoption of EVs in the current and future global transportation system, special attention must be given to solve the dynamic charging requests under the real-world complicated situation.

EVs are taking shape as the potential solution for the environmental and energy problems. However, since the limited driving range and insufficient charging infrastructure cause trouble for the EV drivers' travel, it calls for the effective methods to guide EV drivers to select suitable charging stations and routes. For this reason, EVs have received increased interest from the scientific community. In consideration of the limited driving range, several studies have attempted to find the optimal routes for EVs based on the framework of constrained shortest path problem (Artmeier et al., 2010; Storandt, 2012; Neaimeh et al., 2013). However, the charging behaviour was not involved in the models. Kobayashi et al. (2011) further considered the impacts of charging behaviour and established a route search method for EVs. In this method, the location of charging station is an influencing factor to select the travel routes, besides the driving range. Wang et al. (2018b) designed a geometry-based algorithm for charging guidance based on the charging request information. The algorithm considered the consistency of direction trend between charging routes and destination. Sweda et al. (2017) proposed two heuristic methods to find the adaptive routing and recharging decisions for EVs. The charging costs were involved in the solution. Besides charging processes, Qin and Zhang (2011) and Said et al. (2013) considered the impacts of queuing time on the charging station selection. The queuing theory was used to optimize the charging guidance. Several studies combined the driving time, charging time and queuing time to discuss the charging and route optimization for EVs (Yang et al., 2013; De Weerdt et al., 2016; Zhang et al., 2018). Furthermore, Wang et al. (2014) incorporated the energy constraints in the travel and proposed the energy-aware routing model for EVs. Cao et al. (2012) and Liu et al. (2014) considered the impacts of charging costs on the charging station selection to investigate the EV charging problems. Yagcitekin and Uzunoglu (2016) developed a smart charging guidance strategy based on the double-layer optimization theory. Sun and Zhou (2016) compared the impacts of different factors on EV charging guidance by using a cost-optimal algorithm. The trade-off between traveling cost and time consumed was obtained to guide drivers for traveling and charging. Wang et al. (2018a) integrated drivers' intention of the choices for traveling and charging. A multi-objective model was established to provide guidance for EV charging. The objectives include the minimization of traveling time, charging costs and energy consumption. Moreover, in view of the environmental effects for EV adoption, many studies focused on the charging guidance methods from the perspective of energy consumption, which aim to determine the energy-efficient routes for EVs under different situations (Wang et al., 2013; Abousleiman and Rawashdeh, 2015; Strehler et al., 2017; Fiori et al., 2018; Fernández, 2018). However, the previous methods for charging guidance are mainly based on the problems in a static road network. In such a road network, the time or energy consumed in each link is constant. Thus, the impacts of traffic condition on driving state are ignored in the solution.

In order to improve the accuracy of charging schemes, Alizadeh et al. (2014) incorporated the time-varying traffic conditions in the traveling and charging problem for EVs. An extended transportation graph was used to find the optimal routes. Yi and Bauer (2018) introduced a stochastic decision making framework to investigate the stochastic effects of traffic condition on energy cost. The primal-dual interior point algorithm was used to construct the optimal paths. Zhang et al. (2016) considered the impacts of traffic condition on driving distance, travel time and energy consumption and proposed a multi-objective routing model. The ant colony optimization algorithm was employed to search the optimal routes. Jafari and Boyles (2017) further incorporated the reliability of routes in the solution for an EV travel problem under the road network with stochastic traffic condition. Daina et al. (2017) explored the EV charging problem by considering the uncertain traffic condition based on the random utility theory. The trade-off among driving distance, charging time and costs for charging selection was analysed. Huber and Bogenberger (2015) utilized the real-time traffic information to investigate the time-varying characteristics of the traffic condition and their impacts on the EV driving state. In addition, several works introduced the network equilibrium theory to explore the optimization models for EV charging and traveling (Jiang et al., 2014; He et al., 2014; Xie and Jiang, 2016; Xu et al., 2017). In such models, the traffic condition would be changed with the number of vehicles in each link. However, most of the existing methods assume that the charging requests of EV drivers are predetermined and overlook their dynamic characteristics. In the real-world situation, the charging requests with variable information may occur at different periods. The charging operating centre could not know the information before receiving the charging requests. Therefore, the previous methods do not have the ability to solve the dynamic charging requests in the real-world complicated situation. Aiming at the dynamic charging requests, Hung and Michailidis (2015) proposed a charging guidance strategy based on the queuing modeling framework, where the charging requests occurred according to a general process during a time period. However, the study did not consider the time-varying characteristics of road network. It assumed that EVs operate with a constant speed in the road network. Moreover, the energy consumption was ignored in the method, which has significant influence on the reachability of charging stations in a road network.

As the number of EVs increases in the transportation system, it is obvious that large-scale EVs will operate on urban road networks in the foreseeable future, which would contribute to multiple charging requests during different periods. In this study, we aim to develop charging guidance strategies for large-scale dynamic charging requests in a time-varying road network. Based on the charging request information, the charging guidance strategies are established from two different perspectives. The EV drivers' travel demands and vehicle balance in charging stations are respectively considered in the proposed strategies. Both the strategies could help EV drivers to select optimal reachable charging stations by considering the time-varying traffic condition on the routes. The performance of the charging guidance strategies in various scenarios is explored by considering their impacts on the operation efficiency of charging stations.

The proposed methods may be used by the operators of charging operating centres to provide EV

drivers with optimal selections of charging stations under different situations or by individual drivers to select optimal charging stations during trips.

The contributions of this study are as follows. Firstly, given the large-scale EV operation situation, the dynamic characteristic of charging requests is investigated. By combining the time-varying road network, a charging guidance problem with dynamic charging requests is formulated. Furthermore, a dynamic recursive equation is developed to explore the change trend of EV number in charging stations. Secondly, based on the charging request information, two charging guidance strategies are established from the perspective of travel demands of EV drivers and operation efficiency of charging stations, respectively. Both the strategies have the ability to ensure the reachability of selected charging stations in a time-varying road network. Lastly, simulation examples are presented to demonstrate the proposed strategies. The performance of the two strategies is compared in different simulation scenarios. The application recommendations in terms of the strategies are discussed based on the simulation results.

The remaining portions of this paper are organised as follows. Section 2 describes the dynamic charging guidance problem and introduces the considerations for charging station selection. Section 3 analyses the selection basis for optimal charging station from two different perspectives and presents the charging guidance strategies. In Section 4, the simulation examples are designed to demonstrate the proposed strategies and further compare their performance. Lastly, Section 5 provides the conclusions and the directions for future research.

## 2. Problem description

During trips, EV drivers often need to recharge their vehicles to reach destinations. When drivers notice that the remaining energy of their vehicles may be insufficient to reach their destinations, they would send charging requests to the charging operating centre. The information of charging requests includes drivers' travel destinations and remaining energy of vehicles. The charging operating centre would help drivers to select suitable charging stations according to the charging request information. However, in the real-world situation, charging requests have significant uncertain characteristic from time dimension. As a matter of fact, the charging requests received in different periods may have different information based on the individual demands of drivers. The charging operating centre could not predict the detailed information of charging requests in advance. Given the situation with large-scale charging requests, the multiple charging requests at the identical periods often have different information, including travel destinations and remaining energy. Notably, in the actual situation, the charging operating centre can obtain the information regarding to the locations of EVs and charging stations by using positioning devices. Therefore, EV drivers do not need to send the location information to the charging operating centre. Meanwhile, in the same position, whether the charging requests occur or their detailed information are not predetermined at different periods. Thus, the charging requests in a road network have significant dynamic characteristic. To represent the dynamic characteristic of charging requests, time is slotted with slots normalized to integral units. Let $\{1,...,t,...,T\}$ denotes the set of time slots, where $T$ is the total number of the time slots. With the identical duration for each time slot, the time horizon increases as $T$ increases. Let $C_i^t$ denotes the charging request that occurs in the node $i$ and at time slot $t$. For each charging request $C_i^t$, let $d_i^t$ and $e_i^t$ denote the travel destination and remaining energy, respectively. The travel destination $d_i^t$ and remaining energy $e_i^t$ from different charging requests $C_i^t$ may vary. The charging operating centre could not understand or predict $C_i^t$ before time slots $t$. The decision-making for all the charging requests needs to be determined based on the traffic condition at corresponding time slots.

In the situation with large-scale EVs, the dynamic characteristic of charging requests is one of the challenges for dealing with the charging requests. Besides, the traffic condition in a road network also affects the EV traveling and charging, because it has significant effects on the driving speed and energy consumption for EVs (Bi et al., 2019). More importantly, the traffic condition in a road network often has time-varying characteristic in the real-world situation due to the influence of environmental factors (Huang et al., 2017). Therefore, the energy and time consumed to traverse the same links may vary at different time slots. To improve the effectiveness of the charging guidance scheme, the time-varying characteristic of a road network should be considered. Combining the road network structure and EV operating characteristics, the time-varying road network is defined as $G=(V,A,\tau_a^t,E_a^t)$, where $V$ and $A$ denote the set of nodes and links, respectively. Notably, in the set $V$, there exist two types of nodes, including normal nodes and charging station nodes. The latter have the ability to charge EVs. For problem formulation, we assume the existence of $m$ normal nodes and $n$ charging station nodes in the set $V$. Moreover, $\tau_a^t$ and $E_a^t$ in the $G$ denote the driving time and energy consumption on the link $a$ at time slot $t$, where $a \in A$. In every time slot, the values of $\tau_a^t$ and $E_a^t$ randomly change within a reasonable range, which reflects the time-varying characteristic of the road network. Note that, the conventional optimization problems with time-varying network often assume that the links hold their state for the duration of a time slot (Neely et al., 2005). For the problem of EV charging guidance, the assumption signifies the constant driving time and energy consumption during a separate time slot. The assumption conforms to the traffic condition characteristic in a real-word road network if the duration for each time slot is relatively short. Therefore, we follow such an assumption in the charging guidance problem for EVs.

Combining the dynamic charging requests and time-varying road network, the charging guidance problem for EVs is formulated. According to the definition of set $V$, there exist $m$ normal nodes and $n$ charging station nodes in a road network. Furthermore, the problem assumes that the charging requests occur in the normal nodes only. The charging station nodes couldn't generate charging requests. Such an assumption is reasonable, because only when drivers hard to search nearby charging stations would they send charging requests to the charging operating centre. Otherwise, they have no need to send charging requests. The purpose of charging guidance strategies is to help EV drivers from normal nodes to select suitable charging station nodes based on specific objectives. Let $x_{ij}^{t}$ denotes the binary decision variable in the problem formulation, which is equal to 1 if the charging request generated in normal node $i\,(i=1,...,m)$ at time slot $t$ is assigned to charging station node $j\,(j=1,...,n)$; otherwise, this variable is 0. For the normal nodes in a road network, in every time slot, all of them have the possibility to generate charging requests. To reflect the dynamic characteristic of charging request occurrence, the possibility of the charging request occurring in node $i$ at each time slot is defined, which is denoted as $\lambda_i$ $(0 \le \lambda_i \le 1)$. The $\lambda_i$ values do not have time-varying characteristic, which are influenced by the node location. Notably, although the node $i$ have a constant possibility $\lambda_i$ for every time slot, the travel destination and remaining energy from the charging requests may vary at different time slot. Moreover, suppose that each normal node can generate at most one charging request during a time slot. The assumption conforms to the characteristic of charging request occurrence in actual situation if the duration for each time slot is relatively short.

When solving the charging requests at each time slot, the first step is to ensure that the remaining energy can support the EVs in reaching the target charging stations. In a time-varying road network, the energy consumption between normal nodes and charging stations nodes may vary at different time slots. Thus, before selecting charging stations, the energy consumption on the routes should be observed and only the reachable charging stations can be considered as the candidate ones, as shown in Fig.1.

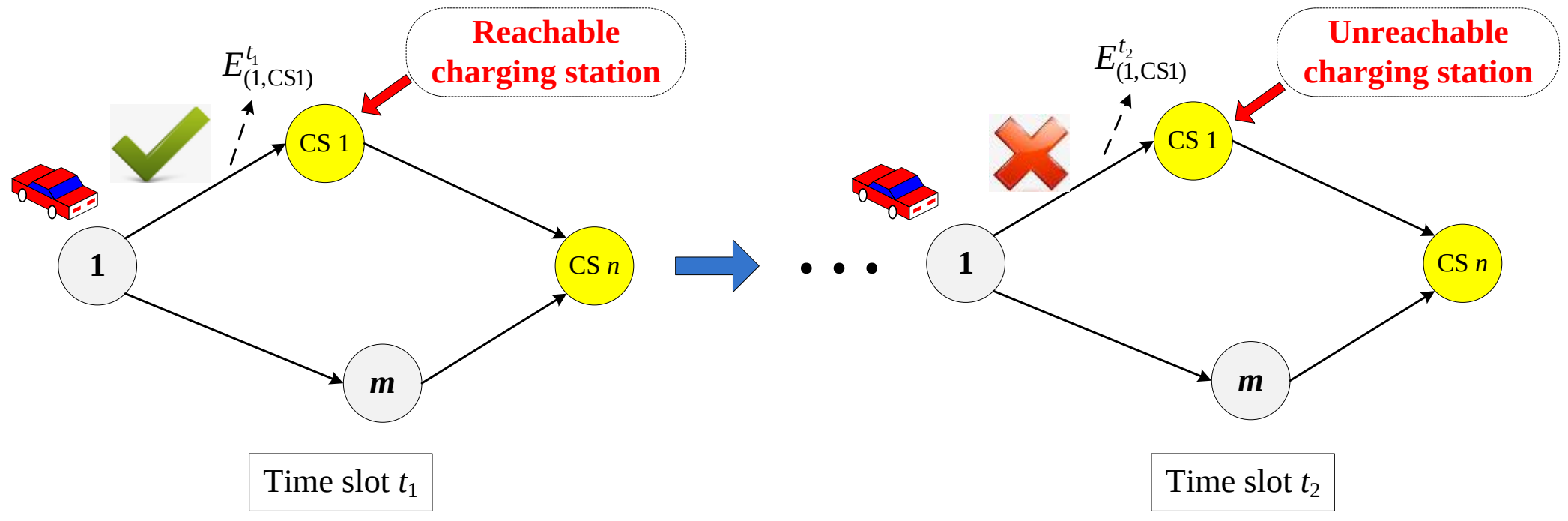

Fig.1 Reachable charging station and unreachable charging station

In Fig.1, the energy consumption between node 1 and CS 1 at time slot $t_1$ and time slot $t_2$ is different. The green check mark represents that the EV have ability to traverse the route. On the contrary, the red cross presents that the EV is unable to traverse the route due to the insufficient remaining energy. The figure indicates that the reachability of the same charging station may vary at different time slots because of the time-varying characteristic of the road network, which is the premise to determine the optimal charging station for each charging request. Furthermore, the driving time on the routes may also change at different time slots, which determines the time slots as the EVs reach charging stations. Assume that the charging operating centre know the information regarding to the traffic condition on all the links at the beginning of each time slot. Such information can be obtained either through real-time traffic information from transport sector or through short-term traffic flow prediction (Polson and Sokolov, 2017). The basic framework for the dynamic charging guidance problem in a time-varying road network is presented in Fig. 2.

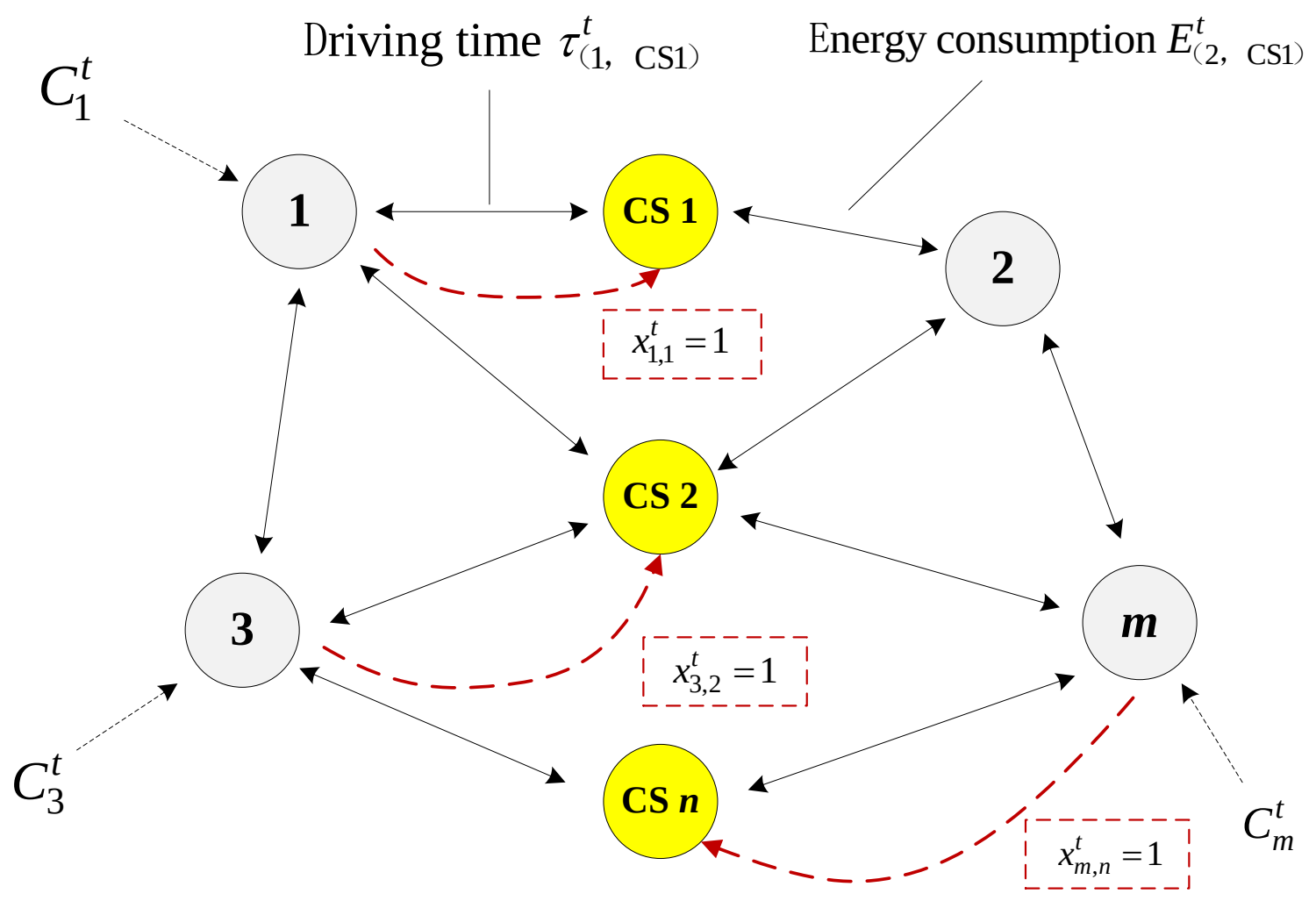

Fig.2 Dynamic charging guidance problem in a time-varying road network

In Fig.2, $\tau^t_{(1,\ CS1)}$ is the driving time from node 1 to charging station node CS 1 under the traffic condition at time slot *t*. $E^t_{(2,\ CS1)}$ is the energy consumption between node 2 and charging station node CS 1 at time slot *t*. Obviously, the charging requests occur in node 1, node 3 and node *m* at time slot *t*, that of $C^t_1, C^t_3$ and $C^t_m$. The objective of the problem is to provide guidance for every charging request by considering the traffic condition at time slot *t*. The optimal charging station nodes would be selected for the charging requests based on specific charging guidance strategies. In the figure, the decision for charging station selection is denoted as $x^t_{(1,1)}=1$, $x^t_{(3,2)}=1$ and $x^t_{(m,n)}=1$. For instance, $x^t_{(1,1)}=1$ indicates that the charging request occurs in node 1 is assigned to the charging station node CS 1. How to determine the value of $x^t_{ij}$ at each time slot *t* is the critical issue to solve the charging guidance problem in a time-varying road network. This issue should be considered from two aspects. On the one hand, the charging guidance strategies satisfy the charging demands of EV drivers. That is, an EV can reach the selected charging station under the current remaining energy. For this reason, the relationship between remaining energy and traffic condition needs to be considered. On the other hand, the charging behavior has significant impacts on the operating state of charging stations, especially in a large-scale EV situation. In every time slot, multiple charging requests may occur in a road network and the charging stations may have to accept multiple EVs. Given the limited charging rate, the number of EVs in a charging station increases as the time slots pass. However, mass EV charging has significant impacts on the operating state of charging stations, which may prolong queuing time and even present a potential burden on local power systems (Putrus et al., 2009; Fernandez and Roman, 2011; Qian et al., 2011). Therefore, besides drivers' charging demands, the number of EVs in each charging station is also an important factor that needs to be considered in charging guidance strategies.

In order to explore the change trend of EV number in charging stations under the situation with large-scale dynamic charging requests, we attempt to develop a dynamic recursive equation based on the operation characteristics of charging stations in a time-varying road network. Let $S_j^t$ denote the number of EVs that complete charging and leave charging station *j* at time slot *t*. Without loss of generality, the problem assumes that at most one EV can leave a charging station at each time slot. The assumption conforms to the actual operating situation of charging stations if the duration for each time slot is relatively short. Moreover, to represent the dynamic characteristic for the EVs leaving charging stations, the possibility of that in charging station node *j* at each time slot is defined and denoted as $\mu_j$ $(0 \le \mu_j \le 1)$. It can be used to reflect the charging levels of the charging stations in a road network. During the actual charging processes, the chargers with different charging levels lead to different charging rates of EVs (Gnann et al., 2018). Notably, under the definition of $S_j^t$ and $\mu_j$, the duration between two adjacent events of an EV leaving charging station *j* follows a geometric distribution (Li and Eryilmaz, 2014). Let $U_j^t$ denote the number of EVs in charging station *j* at time slot *t*. The dynamic recursive equation for $U_j^t$ is

$$U_j^t = \begin{cases} \varphi_j \ , & t=1, j \in \{1,\cdots,n\} \\ \max\{U_j^{t-1} + \sum_{i=1}^{m}\sum_{t''=1}^{t-1} x_{ij}^{t''} - S_j^{t-1}, 0\}, & t \in \{2,\cdots T\}, j \in \{1,\cdots,n\} \end{cases} \tag{1}$$

Where $\varphi_j$ is the initial number of EVs in charging station *j* within the time horizon; $t''$ is the time slot when the charging request from node *i* occurs. In the equation, the time slots $t''$ and $t$ satisfies the following relationship.

$$t = t'' + \tau_{(i,j)}^{t''} \tag{2}$$

Eq. (2) indicates that the EV with charging request $C_i^{t''}$ can reach charging station *j* after the driving time $\tau_{(i,j)}^{t''}$.

Notably, for the problem formulation, the probability variables $\lambda_i$ and $\mu_j$ are introduced to simulate the events of charging request occurrence and EV leaving charging stations during the time horizon. However, in the real-word situation, the charging operating centre could receive the charging request information and know the number of vehicles leaving charging stations at the beginning of each time slot. Therefore, the probability variables $\lambda_i$ and $\mu_j$ do not appear in the dynamic recursive equation. Moreover, without loss of generality, the problem assumes that the routes with minimum energy consumption are selected as the travel routes between departure points and charging station nodes. Furthermore, assume that EV drivers can reach their destinations by charging their vehicles only once, because trips with more than one charging are generally uncommon in urban road network (Franke and Krems, 2013).

## 3. Charging guidance strategies for dynamic charging requests in a time-varying road network

To solve the dynamic charging requests in a time-varying road network, the charging station selection decisions at each time slot should be made on the basis of specific strategies. In this section, we attempt to develop the charging guidance strategies from two different perspectives. Firstly, the travel demands of EV drivers are considered in the strategy. Secondly, the other strategy focuses on the impacts of large-scale charging requests on charging stations. The effectiveness and comparison of the two strategies will be discussed in Section 4. Notably, to guarantee the existence of solution, assume that every charging request has at least one reachable charging station in a road network. For the special situation that no reachable charging station exists, the extra cost may occur to transport the vehicles, such as the trailer service, which is not discussed in the charging guidance strategies.

### *3.1 The charging guidance strategy based on travel demands of EV drivers*

EV drivers are the decision makers for travel activities and also the service objectives of smart charging service. Therefore, it is necessary to consider the travel demands of EV drivers when planning the selection strategy of charging stations. As expected, the reachability is the most critical factor for charging station selection in an EV trip. On the premise of charging station reachability, EV drivers often desire to reduce their travel cost as much as possible. In general, the travel cost is regarded as the optimization criterion to choose the travel routes (Gao et al., 2010; Braekers et al., 2015). The travel cost minimization is one of the critical factors for travel demands. For an EV trip, the travel cost has multi-dimensional components, such as travel time, energy consumption and charging cost (Wang et al., 2018a). Specifically, both the driving time and energy consumption have close correlation to the driving distance. Generally, the driving time is typically proportional to the driving distance with a constant driving speed. Furthermore, the energy consumption also has the significant linear relationship with driving distance (Bi et al., 2018b). Since the energy consumption has significant influence on charging cost, the charging cost would be affected by the driving distance. Thus, the driving distance can be used to reflect the integration of travel cost components. For this reason, the driving distance minimization is employed to establish the charging guidance strategy based on drivers' travel demands.

Note that, different from the driving time and energy consumption, the driving distance is a static factor in a time-varying road network. Adopting driving distance as selection criterion can utilize such an advantage and avoid the complicated dynamic prediction. Moreover, in the charging guidance strategy, the driving distance from charging stations to destinations is considered. For the routes from origins to charging stations, EV drivers prefer to focus on the reachability rather than distance. Thus, the driving distance between origins and charging stations is not involved in the strategy. To simplify the description, the charging guidance strategy based on travel demands of EV drivers is represented by SDD (shortest driving distance) strategy. By using a toy road network with four nodes, Fig. 3 presents the selection basis for optimal charging station under SDD strategy.

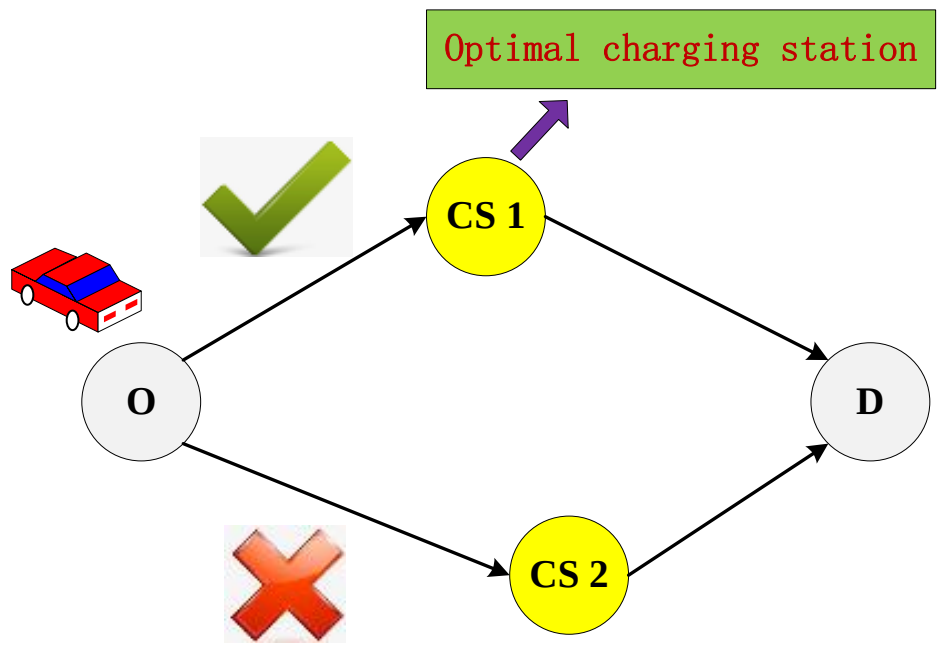

(1) One reachable charging station

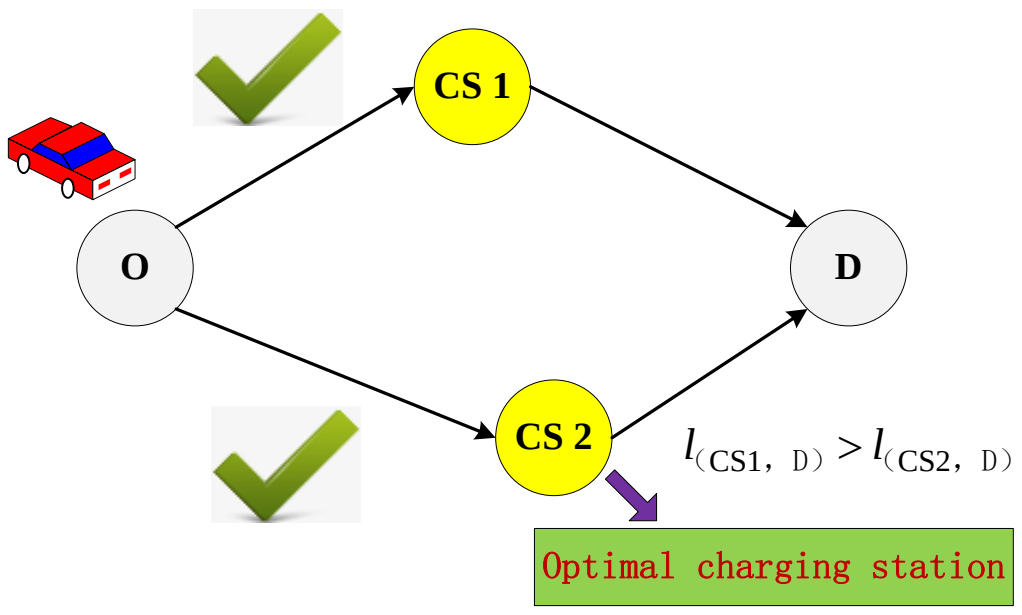

(2) Two reachable charging stations

Fig.3 Selection basis for optimal charging station under SDD strategy

As shown in Fig.3, there exist two charging stations in the toy road network, that of CS 1 and CS 2. The node O and D represent the origin and destination for the EV with charging request, respectively. Furthermore, the figure is divided into two cases according to the number of reachable charging stations. In the case (1), CS 1 is the reachable charging station but CS 2 is the unreachable one. In this case, CS 1 is the optimal charging station because only CS 1 can be selected to charge the EV. In the case (2), both CS 1 and CS 2 are reachable for the EV. In such a case, the driving distance between node D and charging stations should be further considered. Let $l_{(CS2,D)}$ denote the distance between CS 2 and node D. In the case (2), suppose $l_{(CS2,D)}$ is shorter than $l_{(CS1,D)}$. Therefore, CS 2 is selected as the optimal charging station because the shorter driving distance contributes to less travel cost.

As mentioned in Section 2, in the dynamic charging guidance problem, the charging operating centre would receive charging requests in every time slot, that of $C_i^t \triangleq \{d_i^t, e_i^t\}$. The energy and time consumed to traverse each link are known at the beginning of each time slot, that of $E_a^t$ and $\tau_a^t$.The SDD strategy aims to determine optimal charging stations, driving time and corresponding routes for all the charging requests at every time slot. Based on the selection basis as shown in Fig.3, the operating steps of SDD strategy are detailed as follows:

*Step 1*: At the beginning of time slot *t*, based on the information of $E_a^t$, calculate the minimum energy consumption between all the nodes with $C_i^t$ and charging station nodes by using the shortest path algorithms (Fu et al., 2006). The minimum energy consumption between charging request node *i* and charging station node *j* is denoted as $E_{(i,j)}^t$. Record $E_{(i,j)}^t$ and corresponding routes.

*Step 2*: For each $C_i^t$ at time slot *t*, compare $E_{(i,j)}^t$ with $e_i^t$. If $e_i^t \geq E_{(i,j)}^t$, the charging station *j* is denoted as reachable charging station $j'$. Otherwise, the charging station *j* is regarded as unreachable one and deleted from the candidate charging stations.

*Step 3*: For all the reachable charging station $j'$ of $C_i^t$, calculate the shortest driving distance between $d_i^t$ and the node with charging station $j'$. The results are denoted as $l_{(j',d_i^t)}$.

*Step 4*: For each $C_i^t$, compare the shortest driving distance between $d_i^t$ and reachable charging station $j'$. Let $j*$ denote the node with optimal charging station. The minimum driving distance between $d_i^t$ and optimal charging station $j'$ needs to satisfy the following condition:

$$l_{(j*,d_i^t)} = \min_{j'}\left\{l_{(j',d_i^t)}\right\} \tag{3}$$

For the decision variable $x_{ij}^t$, its values can be determined as follows:

$$x_{ij}^t = \begin{cases} 1, & j = j^* \\ 0, & j \neq j^* \end{cases} \tag{4}$$

Moreover, if there exist multiple charging stations with same and minimum driving distance between them and $d_i^t$, randomly select one as the optimal charging station for $C_i^t$.

*Step 5*: Calculate and record the driving time on the minimum energy routes between nodes with $C_i^t$ and corresponding optimal charging station nodes $j*$. The results are denoted as $\tau_{(i,j*)}^t$.

*Step 6*: Before the end of time slot *t*, output the optimal charging stations $j*$, driving time $\tau_{(i,j*)}^t$ and corresponding routes for each charging request $C_i^t$.

Fig. 4 presents the flowchart of SDD strategy.

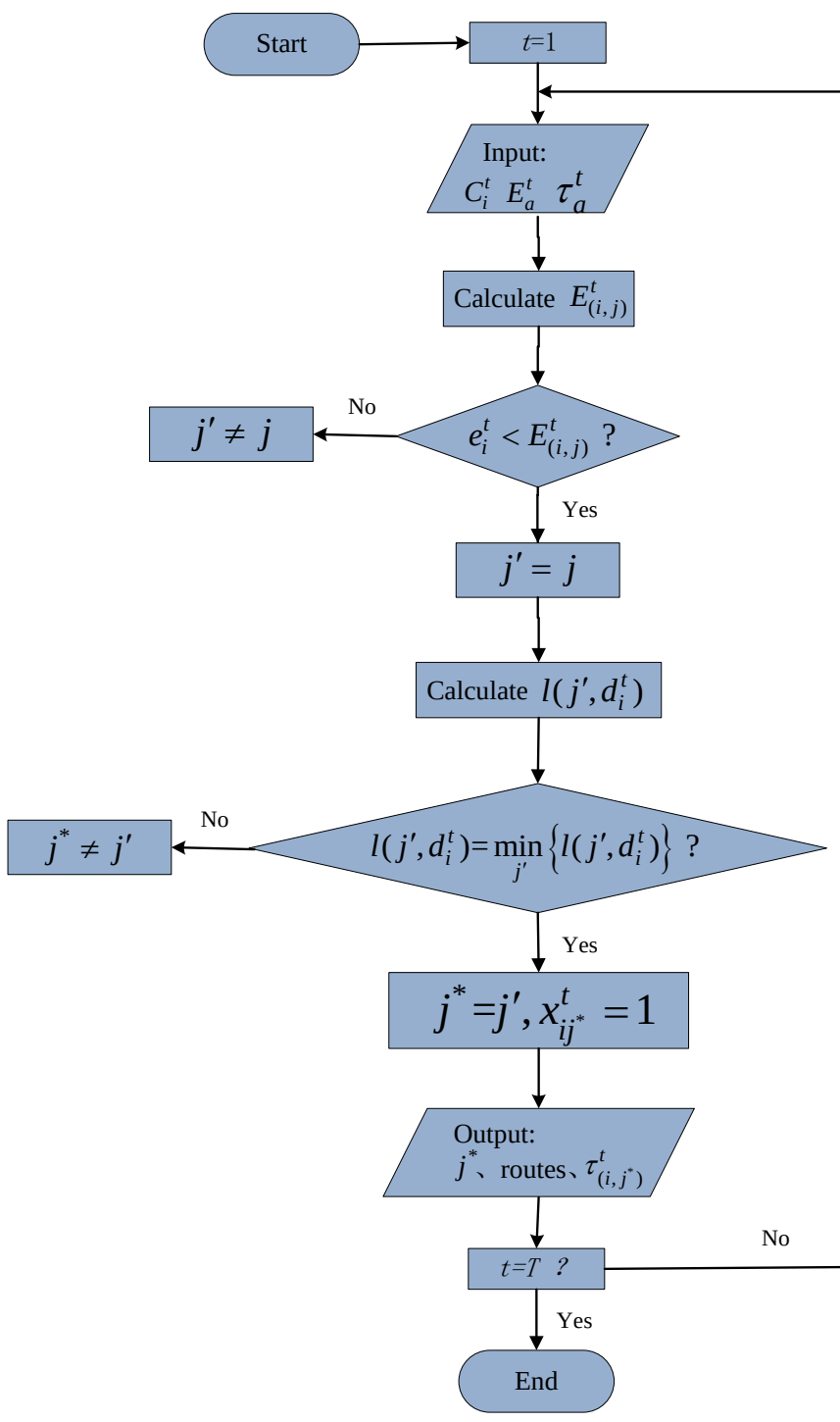

Fig.4 Flowchart of SDD strategy.

*3.2 The charging guidance strategy based on vehicle balance in charging stations*

In the real-world traveling situation, especially with large-scale EVs, a lot of charging requests may occur in the road network. In every time slot, the optimal charging stations need to be selected for all the charging requests. However, as compared to the increasing number of EVs, the number of charging infrastructure is often insufficient in the transportation system. Given the limited charging technology in current and foreseeable future periods (Raslavičius et al., 2015; Bi et al., 2018a), the large-scale charging requests and insufficient charging infrastructure may lead to queuing processes in charging stations. It is noted that, the increasing number of EVs in charging stations has significant impacts on the charging station operation. On the one hand, mass EV charging may increase the operating burden of charging stations and power systems. On the other hand, the queuing time in charging stations may increase as vehicle number increases, which significantly affects the service level of charging guidance. Therefore, to ensure the operation efficiency of charging stations and service satisfaction of drivers, the number of vehicles in each charging station should be considered when solving the large-scale charging requests.

Facing large-scale charging requests, in order to ensure all charging stations with stable operation state, it is useful to select charging stations with relatively small number of EVs for each charging request. However, in the actual situation, the chance of being selected for charging stations has significant difference if the EV number in charging stations is overlooked. For example, the charging stations located in centre areas may accept more EVs with charging requests than other ones. Thus, neglecting the number of EVs in charging stations would enlarge the number of EVs in the charging stations that are located in centre areas. For a transportation system, the charging service is stable if the state of all charging stations is stable. Therefore, balancing the vehicle number in different charging stations is an effective method to alleviate the negative influence of large-scale EVs with dynamic charging requests. For this reason, the charging guidance strategy based on vehicle balance in charging stations is established. On the premise of charging station reachability, the strategy aims to select the charging stations with minimum vehicle number as the optimal ones for the charging requests at each time slot. To simplify the description, the charging guidance strategy based on vehicle balance in charging stations is represented by CSB (charging station balance) strategy. The selection basis for optimal charging station under CSB strategy is presented in Fig. 5.

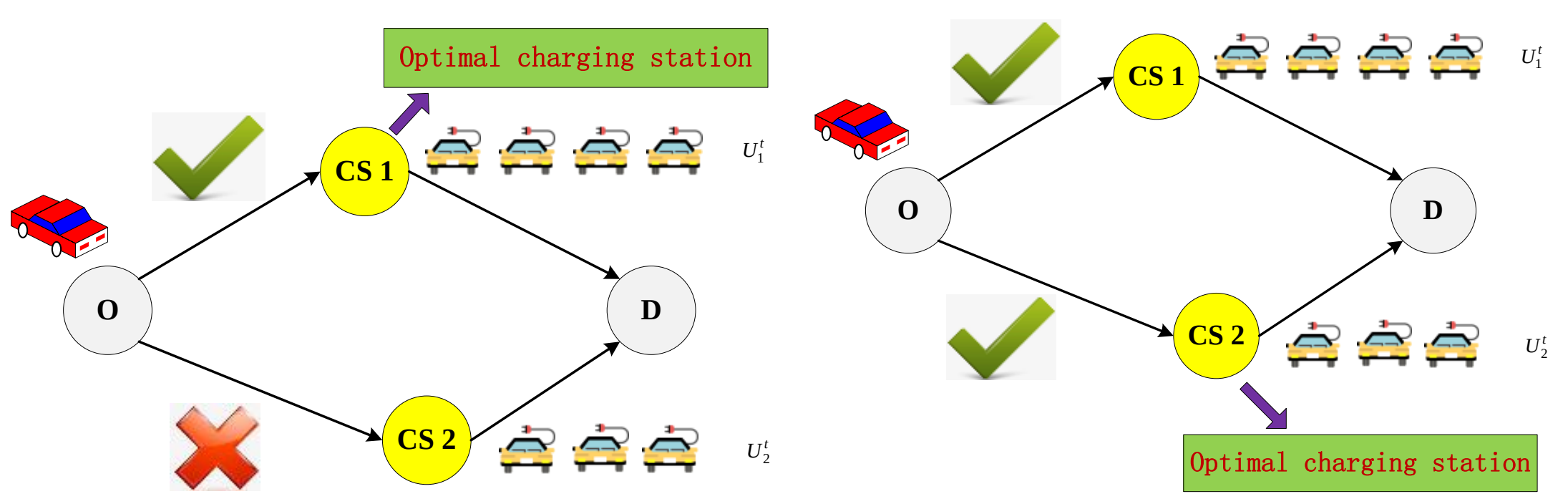

(1) One reachable charging station (2) Two reachable charging stations

Fig.5 Selection basis for optimal charging station under CSB strategy

Fig.5 is divided into two cases based on the number of reachable charging stations. Moreover, the number of EVs in CS 1 is larger than that in CS2. In the case (1), CS 1 is the optimal charging station because only CS 1 is the reachable charging station for the EV. In the case (2), the vehicle number in the charging stations should be considered because both the charging stations are reachable. As shown in the figure, in such a case, CS 2 is selected as the optimal charging station because the number of EVs in CS 2 is less than that of CS 1.

Similar with the SDD strategy, the CSB strategy also aims to determine optimal charging stations, driving time and corresponding routes at every time slot, based on the information from each charging request. The operating steps of CSB strategy are detailed as follows:

*Step 1*: At the beginning of time slot $t$, based on the information of $E_a^t$, calculate the minimum energy consumption between all the nodes with $C_i^t$ and charging station nodes. The minimum energy consumption between charging request node $i$ and charging station node $j$ is denoted as $E_{(i,j)}^t$. Record $E_{(i,j)}^t$ and corresponding routes.

*Step 2*: For each $C_i^t$ at time slot *t*, compare $E_{(i,j)}^t$ with $e_i^t$. If $e_i^t \geq E_{(i,j)}^t$, the charging station *j* is denoted as reachable charging station $j'$. Otherwise, the charging station *j* is regarded as unreachable one and deleted from the candidate charging stations.

*Step 3*: For all the reachable charging station $j'$ of $C_i^t$, check the EV number in charging station $j'$ at time slot *t*. The results are denoted as $U_{j'}^t$.

*Step 4*: For each $C_i^t$, compare the EV number $U_{j'}^t$ in all the reachable charging stations $j'$ at time slot *t*. Let $j*$ denote the node with optimal charging station. The optimal charging station $j*$ needs to satisfy the following condition:

$$j^* = \arg \min_{j'} \{U_{j'}^t\} \tag{5}$$

For the decision variable $x_{ij}^t$, its values can be determined on the basis of Eq. (4). Moreover, if there exist multiple charging stations with same and minimum EV number, randomly select one as the optimal charging station for $C_i^t$.

*Step 5*: Calculate and record the driving time on the minimum energy routes between nodes with $C_i^t$ and corresponding optimal charging station nodes *j**. The results are denoted as $\tau_{(i,j*)}^t$.

*Step 6*: Before the end of time slot *t*, output the optimal charging stations *j**, driving time $\tau_{(i,j*)}^t$ and corresponding routes for each charging request $C_i^t$.

Fig. 6 presents the flowchart of CSB strategy.

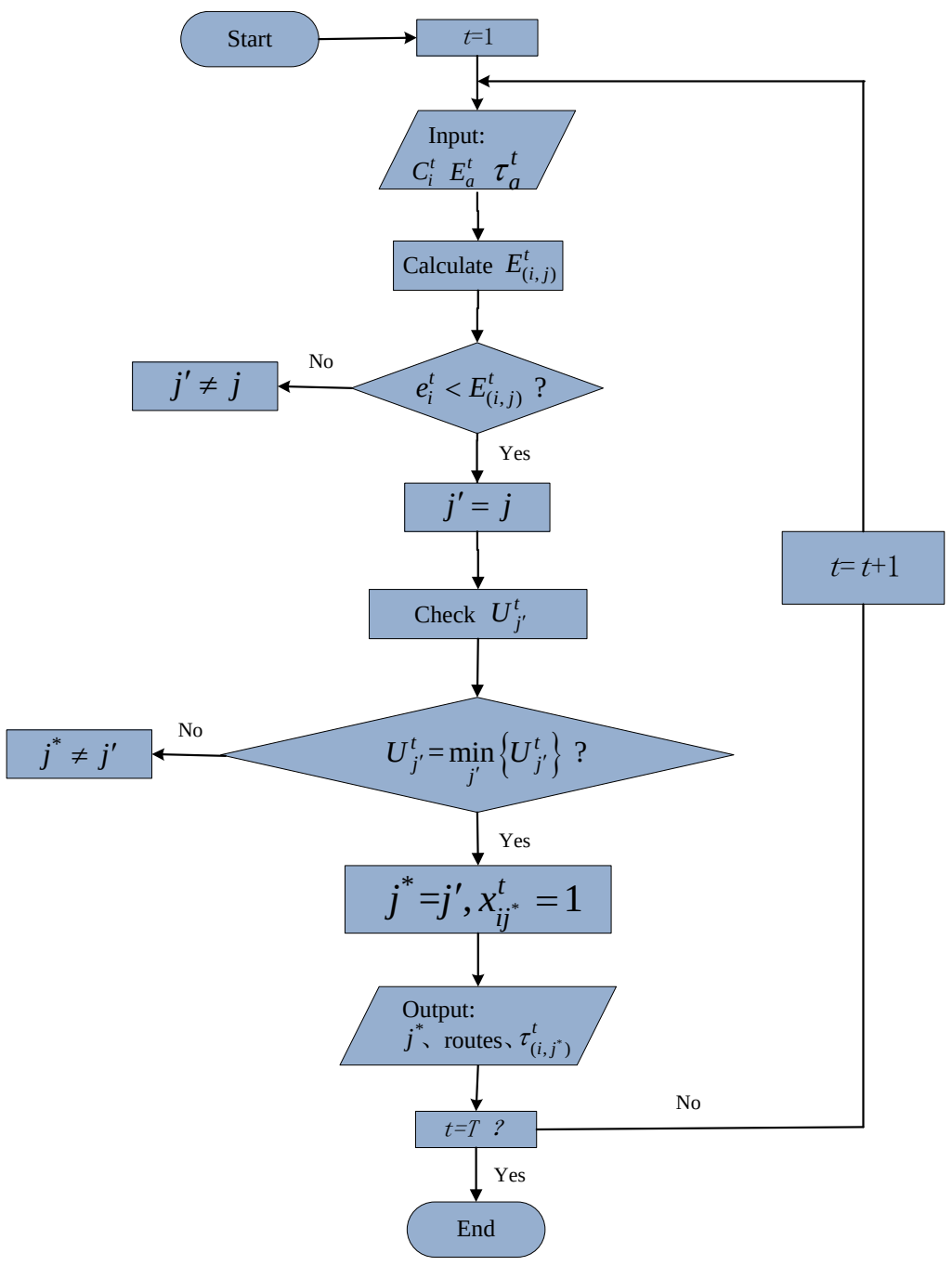

Fig.6 Flowchart of CSB strategy.

## 4. Simulation example

*4.1 Example scenario description*

In this section, we present a simulation example to demonstrate the proposed charging guidance strategies. A time-varying road network is introduced to implement both the SDD and CSB strategies. The structure of the road network is designed based on the Sioux Falls network, which is often adopted to simulate travel optimization problems (Meng and Yang, 2002; Chow and Regan, 2011; Bell et al., 2017). The network consists of 24 nodes and 76 links, as shown in Fig. 7. The road network has eight charging stations, and the nodes with charging stations (yellow) are marked as CS 1 to CS 8. The other nodes, numbered as 1 to 16, are the normal ones without charging station, which may generate charging requests in every time slot.

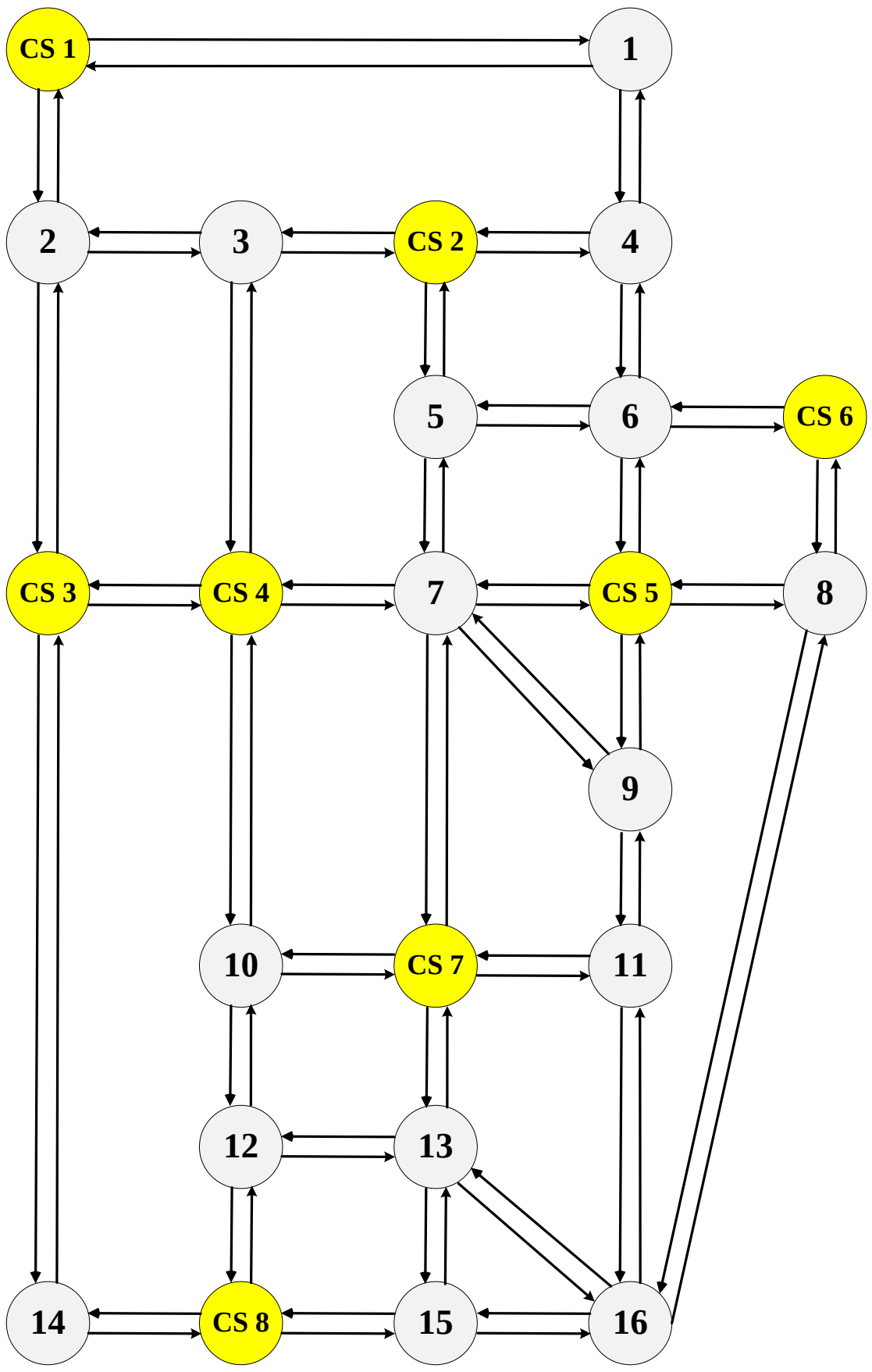

Fig.7 Sioux Falls road network for simulation example.

In the road network, each charging station node has the parameter $\mu_j$ to reflect the charging levels of the charging station, where $j$={1,2,…,7,8}. For the normal nodes, each one has the parameter $\lambda_i$ to reflect the dynamic characteristic of charging request generation, where $i$={1,2,…,15,16}. Table 1 and Table 2 list the value of $\mu_j$ and $\lambda_i$, respectively. Additionally, the simulation example assumes that the initial number of EVs in each charging station equals to zero, that of $\varphi_j = 0$ for all the charging stations $j$ in the road network.

Table 1

Value of parameter $\mu_j$ for each charging station node.

| Charging station | $\mu_j$ | Charging station | $\mu_j$ | Charging station | $\mu_j$ |
|---|---|---|---|---|---|
| CS 1 | 0.74 | CS 4 | 0.90 | CS 7 | 0.94 |
| CS 2 | 0.84 | CS 5 | 0.78 | CS 8 | 0.90 |
| CS 3 | 0.94 | CS 6 | 0.87 | | |

Table 2

Value of parameter $\lambda_i$ for each normal node.

| Normal node | $\lambda_i$ | Normal node | $\lambda_i$ | Normal node | $\lambda_i$ | Normal node | $\lambda_i$ |
|---|---|---|---|---|---|---|---|
| 1 | 0.31 | 5 | 0.20 | 9 | 0.18 | 13 | 0.57 |
| 2 | 0.62 | 6 | 0.13 | 10 | 0.27 | 14 | 0.35 |
| 3 | 0.32 | 7 | 0.50 | 11 | 0.15 | 15 | 0.52 |
| 4 | 0.69 | 8 | 0.25 | 12 | 0.26 | 16 | 0.67 |

For each charging request $C_i^t$, its information includes the travel destination $d_i^t$ and remaining energy $e_i^t$, which is randomly generated in the simulation example. In every time slot, if a charging request occurs in a specific normal node, the travel destination $d_i^t$ is randomly selected from other normal nodes in the road network. The remaining energy $e_i^t$ is randomly obtained from a given interval, which ranges from 7.2 kWh to 16.8 kWh in the simulation example. In a time-varying road network, the energy consumption on each link varies as time slot passes. To reflect such a characteristic, at time slot *t*, the simulation example randomly determines the parameter $E_a^t$ from given intervals for link *a*. The value intervals of the energy consumption $E_a^t$ on each link *a* are listed in Table 3.

Table 3

Value intervals of energy consumption $E_a^t$ for the link *a* at time slot *t*

| Link *a* | $E_a^t$ (kWh) | Link *a* | $E_a^t$ (kWh) | Link *a* | $E_a^t$ (kWh) |
|---|---|---|---|---|---|
| 1–CS1 | [2.64,5.76] | 9–7 | [2.16,4.8] | CS1–2 | [1.68,4.08] |
| 1–4 | [3.6,5.04] | 9–11 | [2.16,5.04] | CS2–3 | [3.6,6.96] |
| 2–CS1 | [2.4,4.8] | 10–CS4 | [2.64,5.52] | CS2–4 | [2.88,6.48] |
| 2–3 | [1.92,4.56] | 10–12 | [1.44,4.08] | CS2–5 | [3.6,4.8] |
| 3–2 | [3.12,4.32] | 10–CS7 | [2.64,4.56] | CS3–2 | [2.88,6] |
| 3–CS4 | [1.44,3.84] | 11–9 | [2.16,3.6] | CS3–CS4 | [2.88,5.28] |
| 3–CS2 | [1.2,4.32] | 11–CS7 | [1.2,4.56] | CS3–14 | [2.16,4.8] |
| 4–1 | [2.16,3.6] | 11–16 | [1.44,3.36] | CS4–3 | [1.2,4.8] |
| 4–CS2 | [2.4,5.28] | 12–10 | [2.16,4.8] | CS4–7 | [2.64,6] |
| 4–6 | [3.12,6.48] | 12–13 | [2.4,4.8] | CS4–10 | [3.12,6.72] |
| 5–CS2 | [2.88,5.28] | 12–CS8 | [1.92,4.8] | CS4–CS3 | [2.88,4.8] |
| 5–6 | [2.4,3.6] | 13–CS7 | [3.36,5.04] | CS5–6 | [1.2,3.12] |
| 5–7 | [1.68,3.6] | 13–12 | [2.64,4.56] | CS5–7 | [1.44,3.36] |
| 6–4 | [1.44.5.04] | 13–15 | [1.68,4.56] | CS5–8 | [3.12,5.04] |
| 6–5 | [2.16,3.6] | 13–16 | [1.44,5.04] | CS5–9 | [2.4,5.04] |
| 6–CS5 | [2.16,5.28] | 14–CS3 | [3.12,5.04] | CS6–6 | [2.16,4.08] |
| 6–CS6 | [3.6,5.52] | 14–CS8 | [2.88,6.24] | CS6–8 | [1.92,5.52] |
| 7–5 | [3.6,5.52] | 15–13 | [1.92,5.28] | CS7–7 | [1.68,4.56] |
| 7–CS4 | [1.92,4.32] | 15–CS8 | [1.2,4.56] | CS7–10 | [3.6,6.72] |
| 7–CS7 | [3.36,5.52] | 15–16 | [2.16,5.28] | CS7–11 | [2.16,3.84] |
| 7–9 | [2.64,6.24] | 16–8 | [2.4,4.08] | CS7–13 | [1.44,2.88] |
| 7–CS5 | [2.4,4.32] | 16–11 | [2.4,6] | CS8–12 | [2.4,3.6] |
| 8–CS6 | [3.36,4.8] | 16–13 | [3.36,6.72] | CS8–14 | [1.2,3.84] |
| 8–CS5 | [3.12,4.8] | 16–15 | [3.6,6.96] | CS8–15 | [2.64,5.28] |
| 8–16 | [2.4,3.84] | CS1–1 | [2.16,5.28] | 2–CS3 | [2.16,4.32] |
| 9–CS5 | [1.2,4.32] | | | | |

Besides the energy consumption $E_a^t$, the driving time on each link *a* also has time-varying characteristic. Like the parameter $E_a^t$, in every time slot, the values of parameter $\tau_a^t$ are randomly determined based on given intervals for each link *a*. The value intervals of the driving time $\tau_a^t$ on each link *a* are listed in Table 4. Notably, since the time is slotted into the time slots with identical duration, the number of time slots is used to represent the driving time on each link. Without loss of generality, the duration for each time slot is not constrained in the simulation example. In the real-world situation, the duration for time slots could be valued according to actual requirement.

Table 4

Value intervals of driving time $\tau_a^t$ for the link *a* at time slot *t*

| Link *a* | $\tau_a^t$ | Link *a* | $\tau_a^t$ | Link *a* | $\tau_a^t$ |
|---|---|---|---|---|---|
| 1–CS1 | [2,5] | 9–7 | [1,2] | CS1–2 | [1,3] |
| 1–4 | [1,4] | 9–11 | [1,2] | CS2–3 | [1,2] |
| 2–CS1 | [2,3] | 10–CS4 | [1,2] | CS2–4 | [2,4] |
| 2–3 | [1,2] | 10–12 | [1,2] | CS2–5 | [1,3] |
| 3–2 | [2,3] | 10–CS7 | [2,4] | CS3–2 | [1,2] |
| 3–CS4 | [2,3] | 11–9 | [1,2] | CS3–CS4 | [1,3] |
| 3–CS2 | [1,4] | 11–CS7 | [2,4] | CS3–14 | [1,2] |
| 4–1 | [1,2] | 11–16 | [1,2] | CS4–3 | [1,2] |
| 4–CS2 | [2,4] | 12–10 | [2,3] | CS4–7 | [1,3] |
| 4–6 | [1,2] | 12–13 | [1,2] | CS4–10 | [1,3] |
| 5–CS2 | [1,2] | 12–CS8 | [1,4] | CS4–CS3 | [1,2] |
| 5–6 | [1,2] | 13–CS7 | [1,3] | CS5–6 | [1,2] |
| 5–7 | [2,3] | 13–12 | [1,2] | CS5–7 | [1,2] |
| 6–4 | [1,2] | 13–15 | [2,3] | CS5–8 | [1,2] |
| 6–5 | [1,2] | 13–16 | [1,2] | CS5–9 | [1,2] |
| 6–CS5 | [1,3] | 14–CS3 | [2,4] | CS6–6 | [1,3] |
| 6–CS6 | [1,2] | 14–CS8 | [2,3] | CS6–8 | [2,3] |
| 7–5 | [1,2] | 15–13 | [2,3] | CS7–7 | [1,3] |
| 7–CS4 | [2,5] | 15–CS8 | [2,5] | CS7–10 | [1,3] |
| 7–CS7 | [2,4] | 15–16 | [1,2] | CS7–11 | [1,2] |
| 7–9 | [1,2] | 16–8 | [1,3] | CS7–13 | [1,2] |
| 7–CS5 | [1,3] | 16–11 | [1,2] | CS8–12 | [2,3] |
| 8–CS6 | [2,5] | 16–13 | [1,3] | CS8–14 | [1,2] |
| 8–CS5 | [2,4] | 16–15 | [2,3] | CS8–15 | [1,2] |
| 8–16 | [1,2] | CS1–1 | [1,2] | 2–CS3 | [1,3] |
| 9–CS5 | [2,3] | | | | |

Furthermore, Table 5 lists the length of each link *a* in the road network. In the table, the length of link *a* is denoted as $l_a$ (km). Notably, considering the structure characteristic of road network, the links with symmetric relation have the same length.

Table 5

Length $l_a$ of link *a* in the road network

| Link *a* | $l_a$ (km) | Link *a* | $l_a$ (km) | Link *a* | $l_a$ (km) |
|---|---|---|---|---|---|
| 1–CS1 | 23 | 9–7 | 15 | CS1–2 | 11 |
| 1–4 | 12 | 9–11 | 10 | CS2–3 | 10 |
| 2–CS1 | 11 | 10–CS4 | 14 | CS2–4 | 10 |
| 2–3 | 10 | 10–12 | 11 | CS2–5 | 11 |
| 3–2 | 10 | 10–CS7 | 12 | CS3–2 | 17 |
| 3–CS4 | 20 | 11–9 | 10 | CS3–CS4 | 12 |
| 3–CS2 | 10 | 11–CS7 | 12 | CS3–14 | 22 |
| 4–1 | 12 | 11–16 | 16 | CS4–3 | 20 |
| 4–CS2 | 10 | 12–10 | 11 | CS4–7 | 10 |
| 4–6 | 12 | 12–13 | 12 | CS4–10 | 14 |
| 5–CS2 | 11 | 12–CS8 | 10 | CS4–CS3 | 12 |
| 5–6 | 10 | 13–CS7 | 11 | CS5–6 | 11 |
| 5–7 | 12 | 13–12 | 12 | CS5–7 | 11 |
| 6–4 | 12 | 13–15 | 10 | CS5–8 | 10 |
| 6–5 | 10 | 13–16 | 16 | CS5–9 | 11 |
| 6–CS5 | 11 | 14–CS3 | 22 | CS6–6 | 12 |
| 6–CS6 | 12 | 14–CS8 | 12 | CS6–8 | 12 |
| 7–5 | 12 | 15–13 | 10 | CS7–7 | 18 |
| 7–CS4 | 10 | 15–CS8 | 11 | CS7–10 | 12 |
| 7–CS7 | 18 | 15–16 | 10 | CS7–11 | 12 |
| 7–9 | 15 | 16–8 | 30 | CS7–13 | 11 |
| 7–CS5 | 11 | 16–11 | 16 | CS8–12 | 10 |
| 8–CS6 | 12 | 16–13 | 16 | CS8–14 | 12 |
| 8–CS5 | 10 | 16–15 | 10 | CS8–15 | 11 |
| 8–16 | 30 | CS1–1 | 23 | 2–CS3 | 17 |
| 9–CS5 | 11 | | | | |

*4.2 Simulation results and analysis*

On the basis of the example scenario, the SDD strategy and CSB strategy are applied in the dynamic charging guidance problem with the time-varying road network. Moreover, in order to analyse the performance during different time horizon, the total number of time slots is respectively set as $T$=$10^2$, $T$=$10^3$, $T$=$10^4$, $T$=$10^5$ and $T$=$10^6$. Note that, both SDD and CSB strategies could ensure the reachability of selected charging stations for the charging requests in every time slot, as mentioned in Section 3. That is to say, the charging demands of EV drivers can be satisfied by both the strategies. Therefore, the simulation example focuses on the impacts of the proposed strategies on charging station operation. The number of EVs in a charging station is a critical factor to reflect the operation state of the charging station. Fig.8 presents the average number of EVs in each charging station during different time horizon $T$ under the proposed strategies.

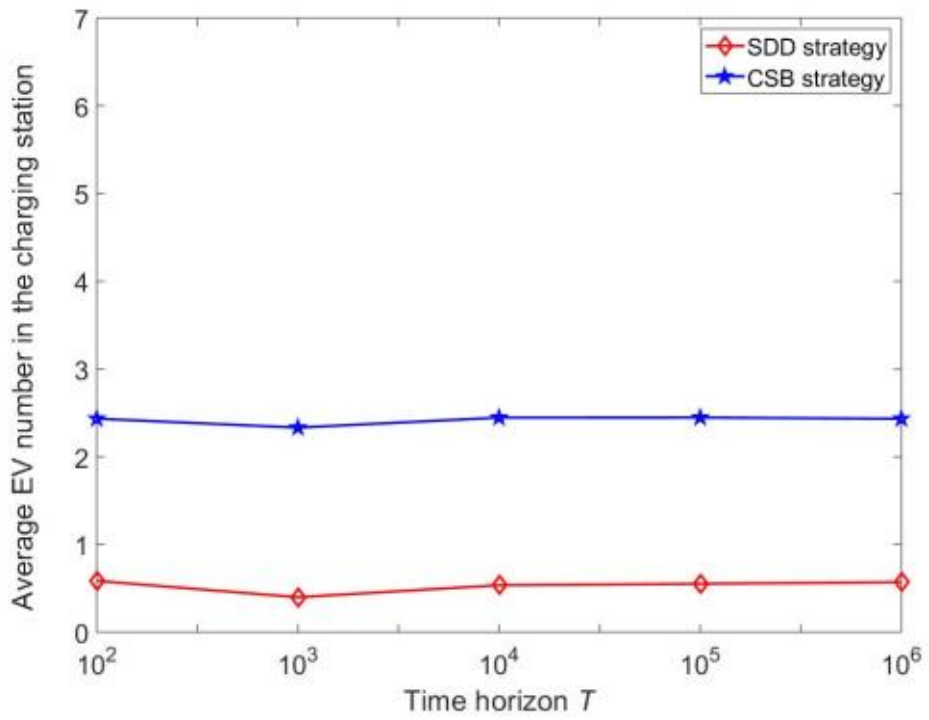

(1) CS 1

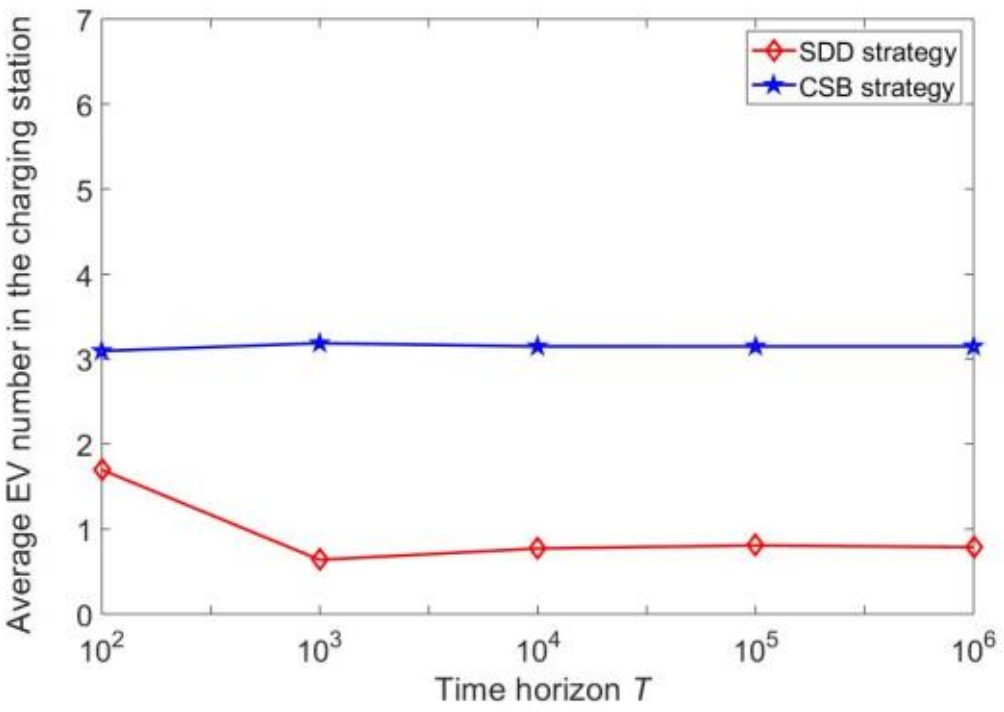

(2) CS 2

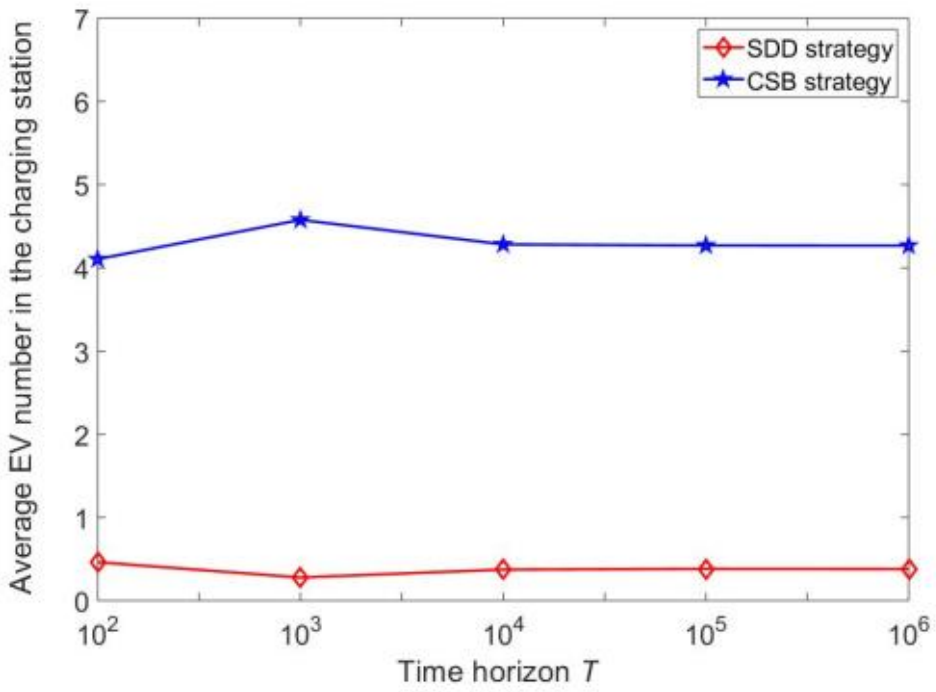

(3) CS 3

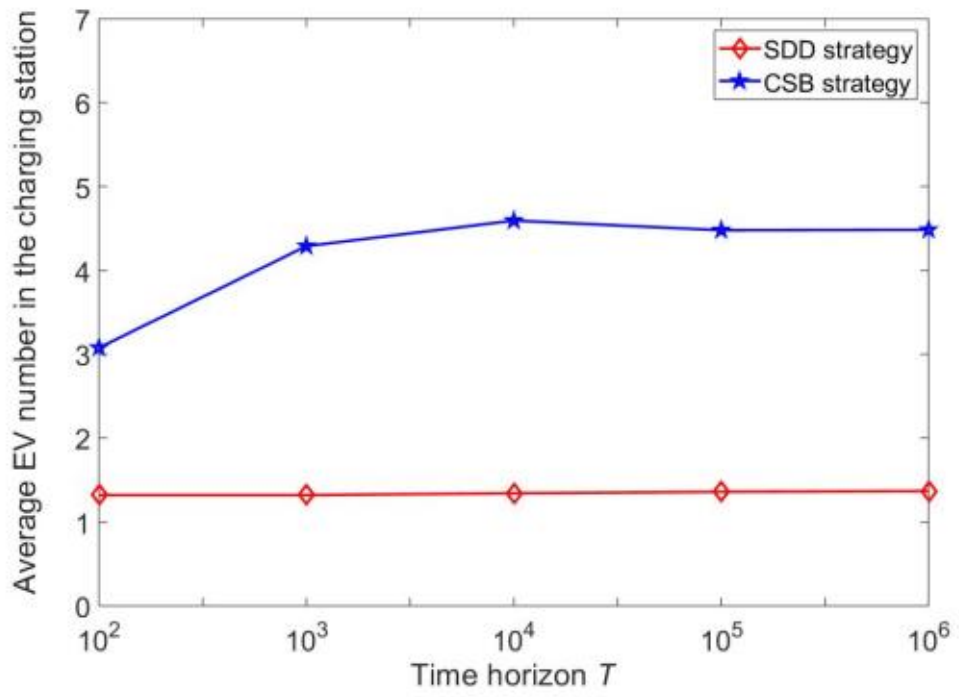

(4) CS 4

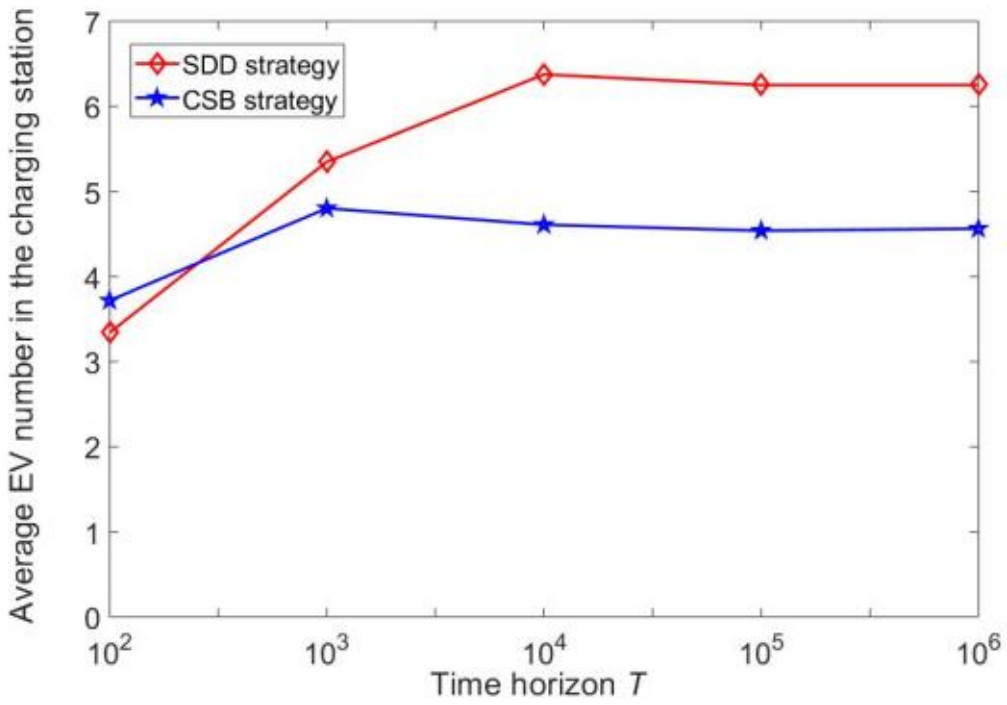

(5) CS 5

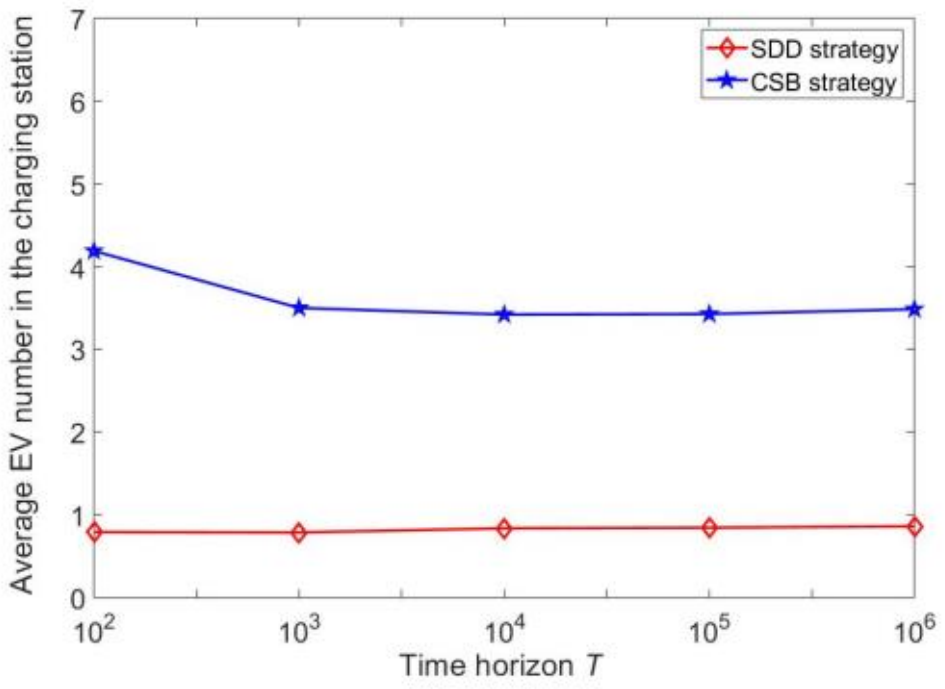

(6) CS 6

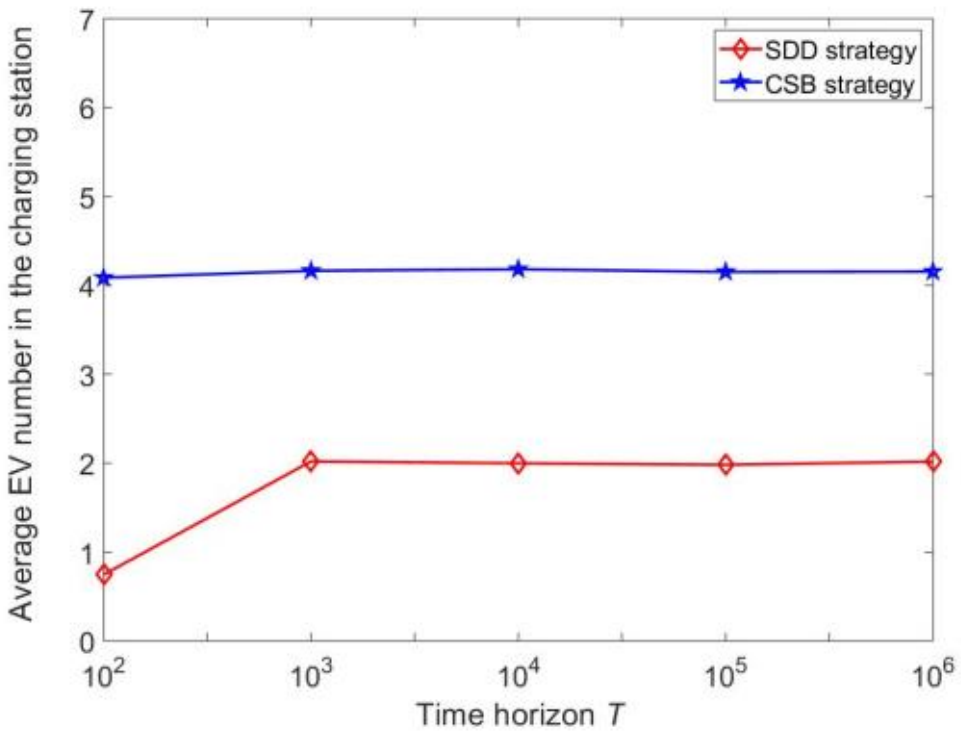

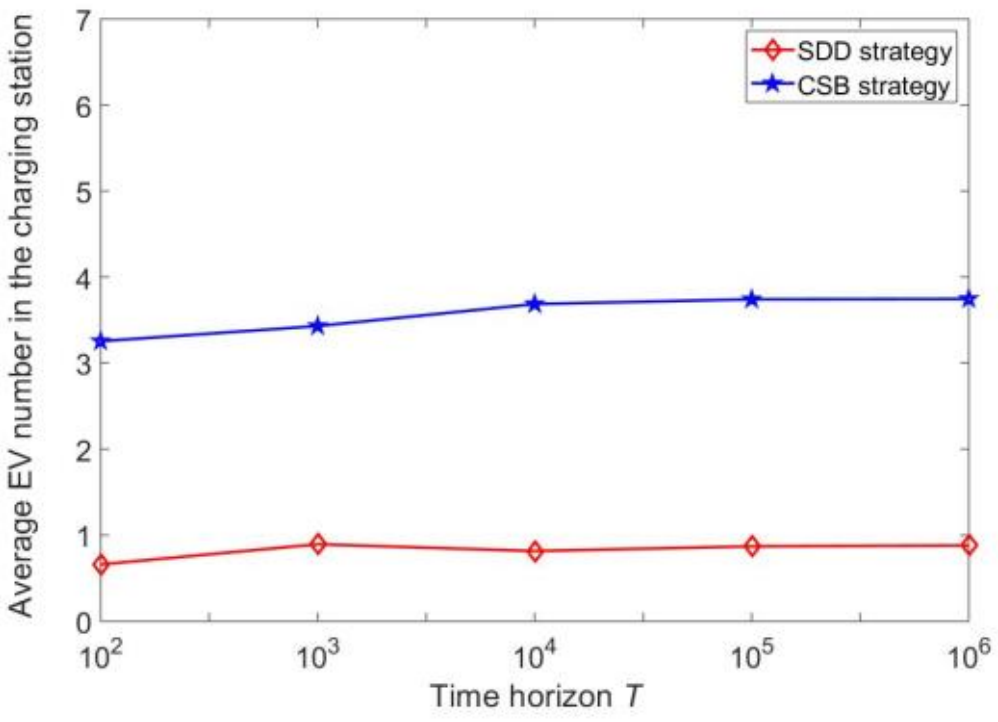

(7) CS 7 (8) CS 8

Fig. 8 Average EV number in each charging station during different time horizon *T*.

In Fig.8, the case (1)-case (8) respectively show the average number of EVs in CS 1-CS 8 during different time horizon. The change trends of the average EV number during different time horizon could reflect the stability of charging stations under specific scenarios. The stability is an important criterion to guarantee the operation efficiency of charging stations. If the average number of EVs in a charging station has a flat change trend as time horizon increases, the charging station would have a stable operation state for the given scenarios (Hung and Michailidis, 2012); otherwise, the average number of EVs in the charging station would have a significant increasing trend as time horizon increases. As can be seen in the figure, both SDD and CSB strategies have ability to stabilise the operation states for CS 1-CS 8 under the example scenario, because the number of EVs in all the charging stations has flat change trends as time horizon *T* varies. Note that, although the fluctuation trends exist when the time horizon ranging from $T=10^2$ to $T=10^4$ for some charging stations under specific strategies, such as the CS 2 under SDD strategy, CS 3 under CSB strategy and CS 5 under both strategies, all the charging stations could reach the stable state after the time horizon $T=10^4$. Moreover, by comparing the average EV number in CS 1-CS 8 with stable state, a significant difference can be observed from the SDD and CSB strategies. For the SDD strategy, the average number of EVs in CS 5 is significant greater than that in other charging stations due to the lack of consideration in vehicle balance of charging stations. On the contrary, the average EV number under the CSB strategy has a similar trend in all the charging stations.

Notably, although the average EV number is a critical reflection of the stability for each charging station, it cannot perfectly represent the actual number of EVs in every time slot. As a matter of facts, the EV number in charging stations at different time slots may varies during the time horizon. Moreover, as time slots pass, the difference between maximum and minimum number of EVs in a charging station may be increasingly significant. If the EV number in a selected charging station is relatively large, drivers would be reluctant to charge their vehicles by using it at corresponding time slot, which would exert negative influences on the implementation efficiency of charging guidance service. Therefore, during the time horizon *T*, the maximum number of EVs in each charging station is often regarded as the bottleneck in the application of charging guidance strategies under the real-world situation. To further compare the performance of SDD and CSB strategies, based on the simulation example, we present the maximum number of EVs in each charging station during different time horizon *T*, as shown in Fig. 9.

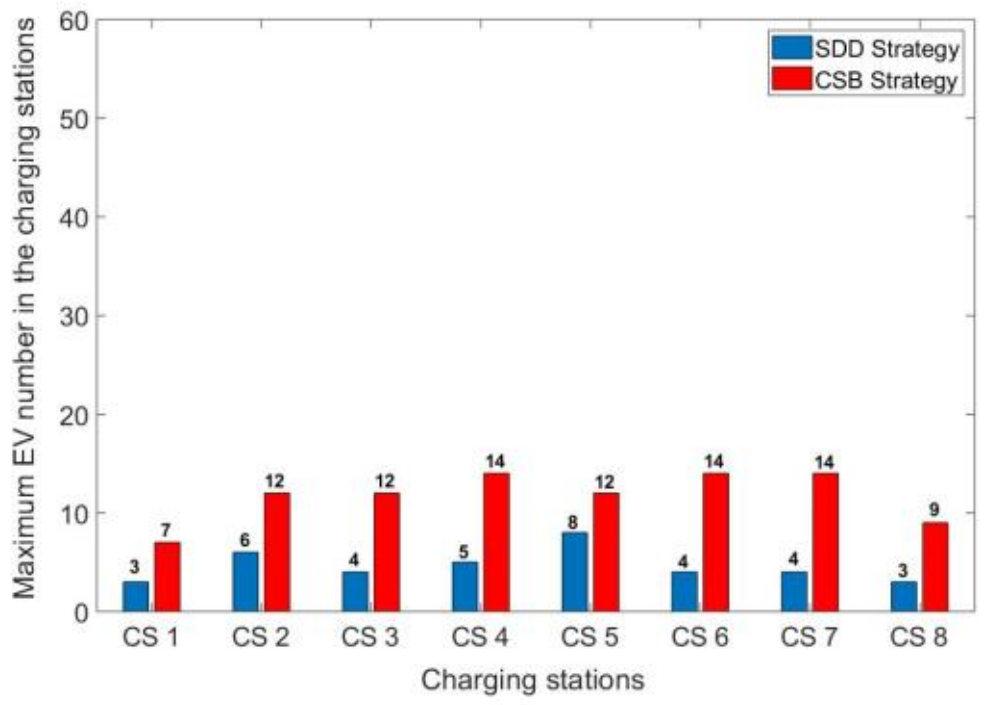

(1) Time horizon $T$=$10^2$

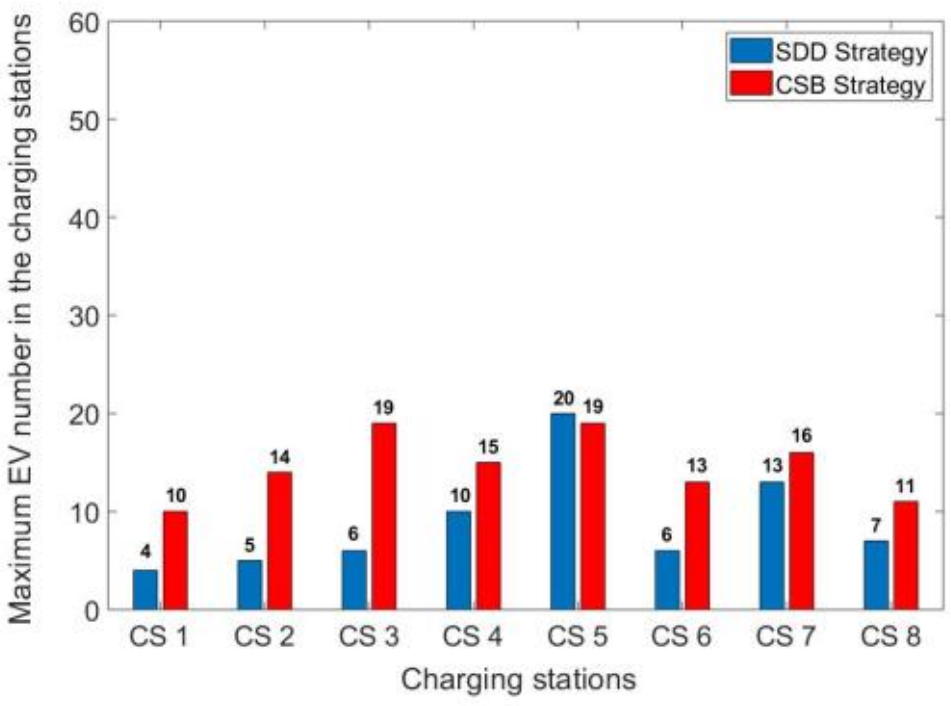

(2) Time horizon $T$=$10^3$

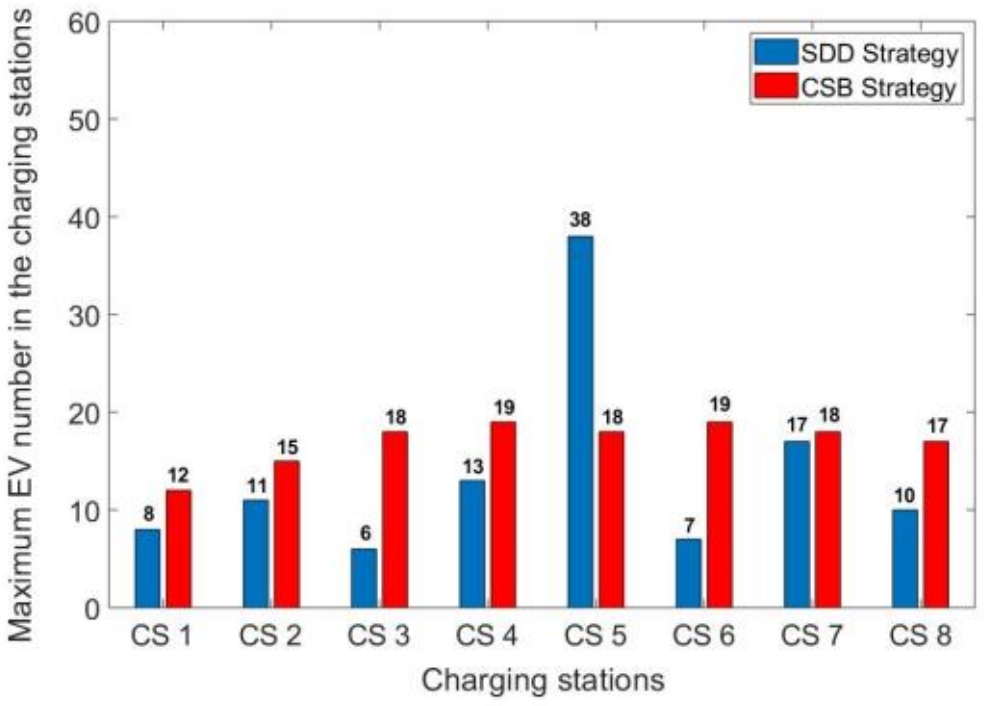

(3) Time horizon $T$=$10^4$

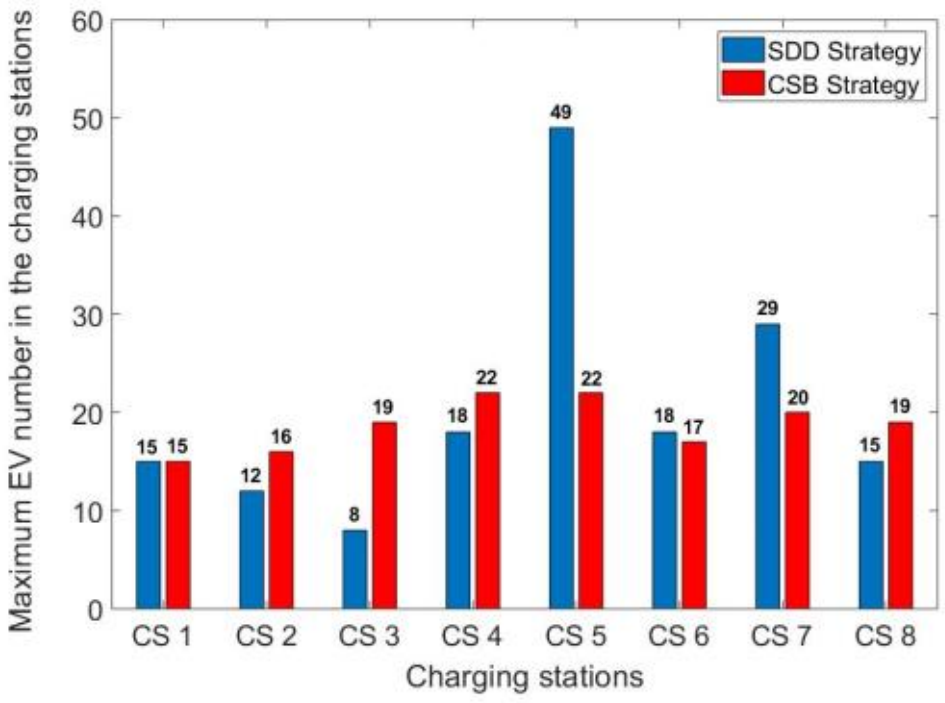

(4) Time horizon $T$=$10^5$

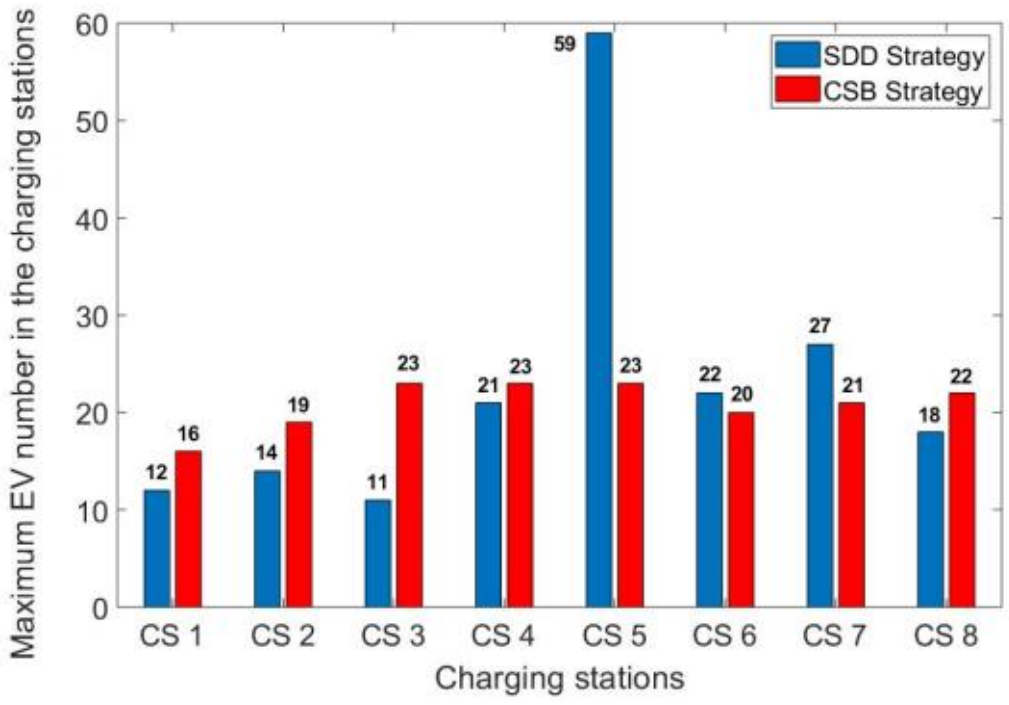

(5) Time horizon $T$=$10^6$

Fig.9 Maximum EV number in each charging station during different time horizon $T$

The maximum number of EVs in each charging station under both SDD and CSB strategies are depicted in Fig.9, where the case (1)-case (5) respectively illustrate the results during different time horizon, ranging from $T$=$10^2$ to $T$=$10^6$. As can be seen, in the case (1), the maximum number of EVs in CS 1-CS8 under the SDD strategy is less than that under the CSB strategy. However, when the time horizon $T$=$10^3$ as shown in the case (2), the EV number in CS 5 under the SDD strategy is larger than that under the CSB strategy. In the case (3), the SDD strategy significantly enlarges the maximum number of EVs in most of charging stations, especially in CS 5, as compared to the case (2). The extreme gap of maximum EV number among the charging stations is equal to 32. On the contrary, the maximum number of EVs under the CSB strategy has a moderate degree of change for all the charging stations. Especially, the maximum number of EVs in CS 3 and CS 5 has a decreasing trend as compared to the case (2). The extreme gap of maximum EV number among the charging stations is equal to 7. When the time horizon $T$=$10^4$ as shown in the case (4), the maximum number of EVs under the SDD strategy increases for all the charging stations, and the maximum EV number in CS 5 and CS 7 is significantly larger than that under the CSB strategy. Moreover, the extreme gap of maximum EV number among the charging stations under SDD and CSB strategies respectively equal to 41 and 7, which presents a significant difference in vehicle balance among different charging stations between the two strategies. In the case (5), the maximum EV number in CS 5, CS 6 and CS 7 under the SDD strategy is larger than that under the CSB strategy. Furthermore, the extreme gap of maximum EV number among the charging stations reaches 48 under the SDD strategy. On the contrary, under the CSB strategy, the maximum number of EVs presents a balanced state for different charging stations. The extreme gap of maximum EV number among the charging stations is equal to 7.

Through comparing the performance of SDD and CSB strategies based on the simulation example, it is observed that the CSB strategy has a significant advantage in terms of the vehicle balance among different charging stations, especially in the situation with long time horizon. Given such a performance, the negative influence resulting from the large number of EVs in a charging station would be avoided by using the CSB strategy. Unlike the CSB strategy, the SDD strategy would enlarge the gap of EV number in different charging stations as time horizon increases, which would affect the operation efficiency of the charging stations that have relatively more vehicles. However, when in the situation with short time horizon, the SDD strategy could be used to deal with the dynamic charging requests, because the difference in performance of the two strategies is not significant in such a situation. More importantly, the travel demands of EV drivers are considered in the SDD strategy.

### *4.3 Parameter analysis*

When discussing the dynamic charging guidance problem, besides time horizon, the scenario characteristics also have significant effects on the performance of proposed strategies. For problem formulation, the parameters $\lambda_i$ and $\mu_j$ are respectively used to present the dynamic characteristics of charging requests and charging processes, as mentioned in Section 2. Note that, such the parameters also have ability to reflect the scenario characteristics in terms of the EV scale and charging level. For instance, a larger parameter $\lambda_i$ represents a larger EV scale in node *i*. Meanwhile, a larger parameter $\mu_j$ illustrates a higher charging level of the charging station in node *j*. In order to explore the performance of charging guidance strategies under different scenario parameters, the parameters $\lambda_i$ and $\mu_j$ are set as different values. Moreover, to highlight the effects of parameter values on simulation results, for each parameter scenario, the values of the parameter $\lambda_i$ for all normal nodes $i$ are set as the identical value $\lambda$. Similarly, the values of parameter $\mu_j$ for all charging station nodes $j$ are set as the identical value $\mu$. Furthermore, the time horizon is set as $T$=$10^6$ for all the parameter scenarios. Fig. 10 presents the maximum number of EVs in each charging station under the SDD strategy, as the parameters $\mu$ and $\lambda$ vary. As can be seen, the value of $\lambda$ is set as 0.1, 0.2, 0.3, 0.4 and 0.5. The value of $\mu$ is set as 0.6, 0.7, 0.8, 0.9 and 1.0. A parameter scenario consists of a pair of parameters $\lambda$ and $\mu$. Thus, totally 25 parameter scenarios are considered.

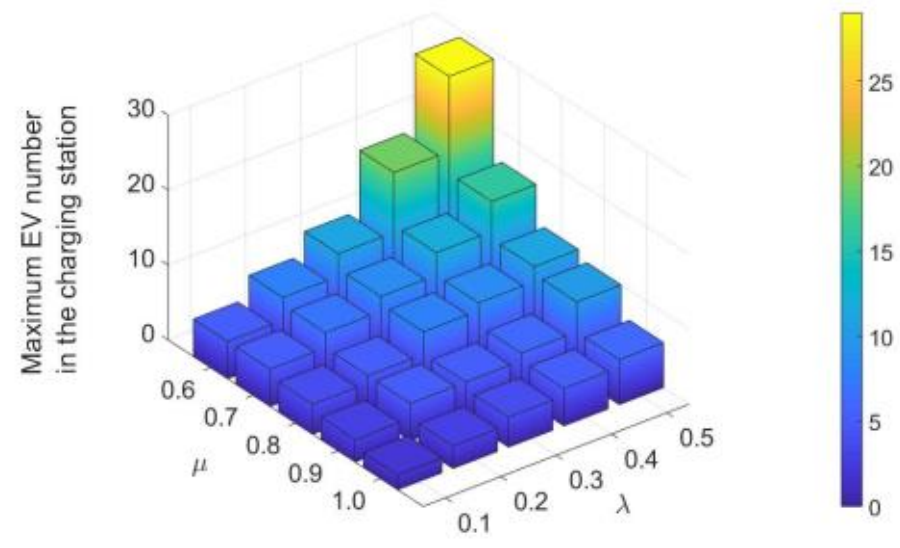

(1) CS 1

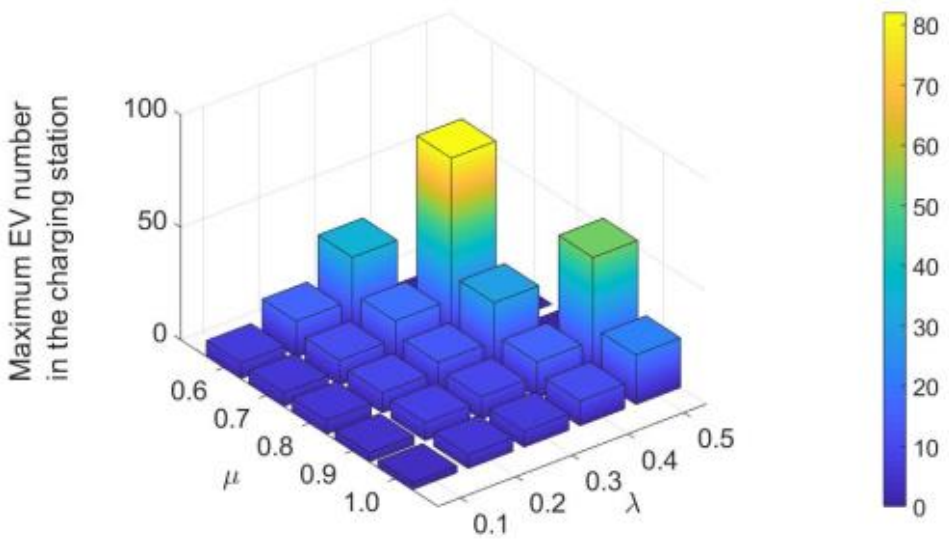

(2) CS 2

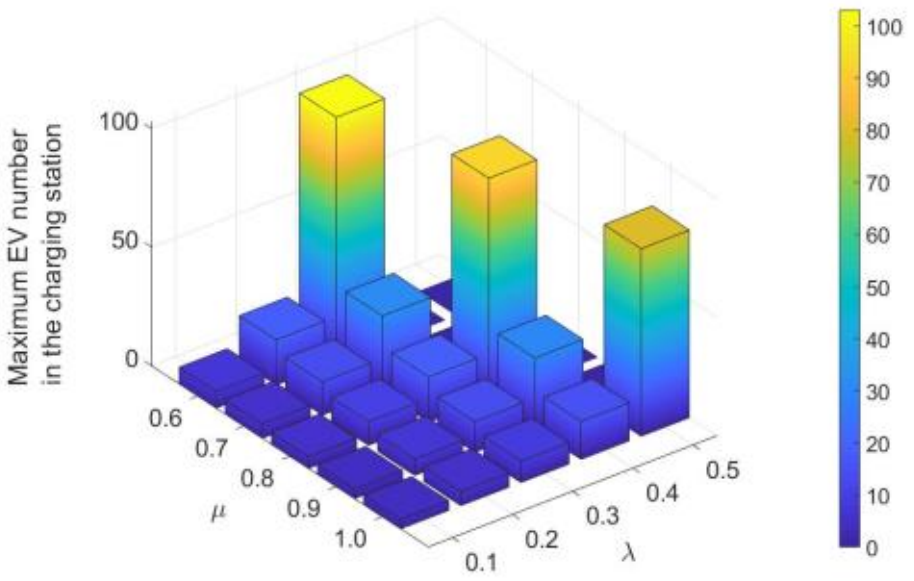

(3) CS 3

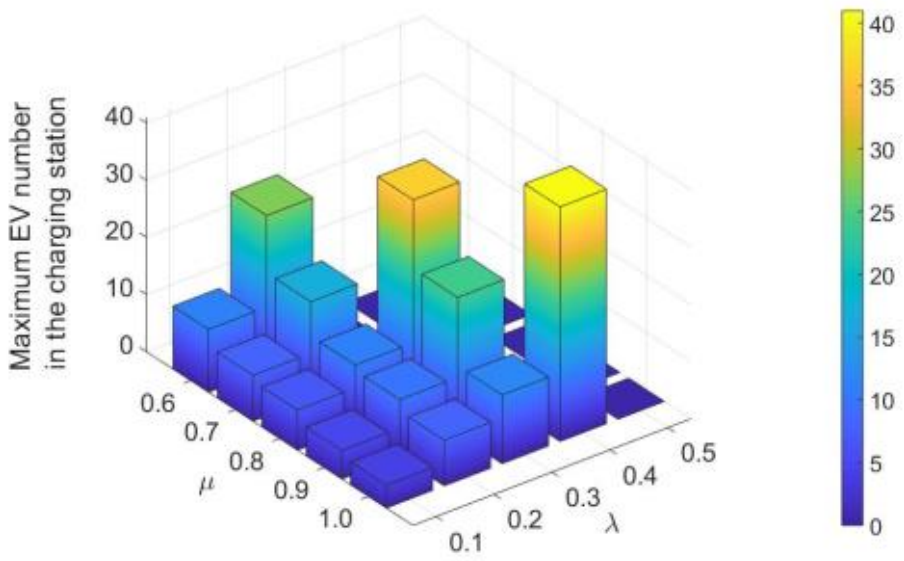

(4) CS 4

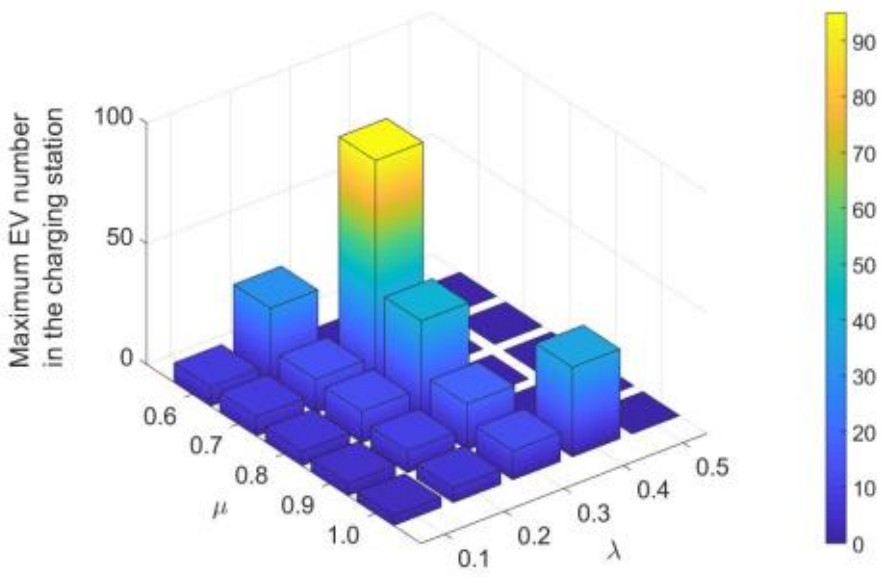

(5) CS 5

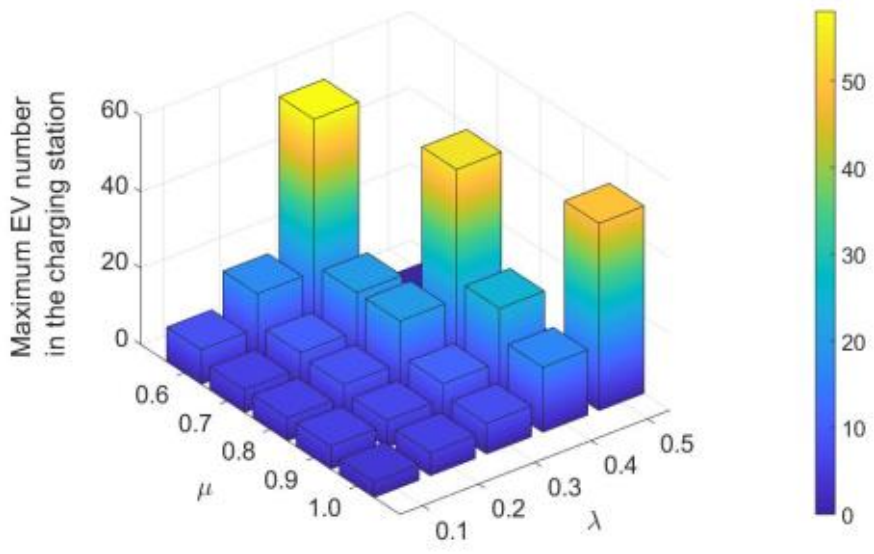

(6) CS 6

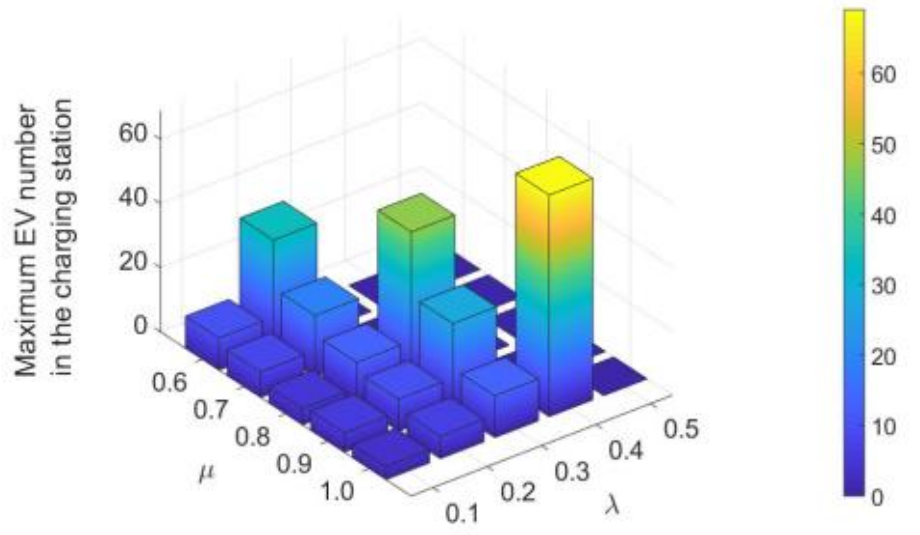

(7) CS 7

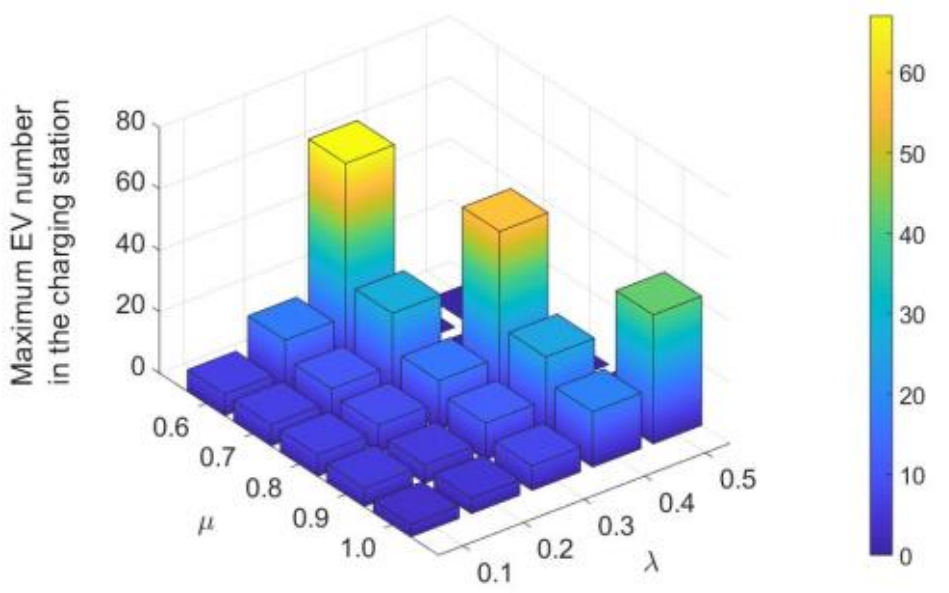

(8) CS 8

Fig.10 Change trends of the maximum EV number in each charging station under the SDD strategy

In Fig.10, the case (1)-case (8) respectively illustrate the maximum EV number in CS 1-CS 8 under the SDD strategy for the different parameter scenarios. Notably, in some parameter scenarios, the maximum EV number in a specific charging station may reach infinity, which indicates that the state of the charging station is unstable. For such a scenario, we let the maximum EV number equals to zero in the figure. Obviously, in the case (1), as the parameter $\mu$ increases, the maximum EV number in CS 1 presents a decreasing trend. This phenomenon indicates that the maximum EV number reduces as the charging level of the charging station increases. In contrast, as the parameter $\lambda$ increases, the maximum number of EVs in CS 1 has an increasing trend, which indicates that the maximum EV number increases as the EV scale increases in the road network. Moreover, among all the parameter scenarios, the peak and lowest values of the maximum EV number are equal to 2 and 29, respectively. In the case (2), as the scenario parameters change, the maximum EV number in CS 2 presents a similar change trend with that in CS 1. However, unlike the case (1), the unstable parameter scenarios exist in the case (2). Among all the stable parameter scenarios, the peak and lowest values of the maximum EV number respectively equal to 3 and 82. Furthermore, in the case (3)-case (8), it is observed that the maximum number of EVs in CS 3-CS 8 also has the similar change trend with that in CS 1. Meanwhile, like CS 2, the unstable state would exist in CS 3-CS 8 under specific parameter scenarios. Among all the stable parameter scenarios, the lowest values of the maximum EV number in CS 3-CS 8 are all equal to 4. Comparatively, the peak values of the maximum EV number in CS 3-CS8 are respectively equal to 103, 41, 95, 58, 69 and 67. Note that, for a transportation system, the charging service is unstable until all the charging stations can reach stability. Therefore, if at least one unstable charging station exists in the road network under a parameter scenario, the SDD strategy cannot be applied in the parameter scenario. Based on such a criterion, the parameter scenarios that cannot support the SDD strategy could be determined. The parameter pairs of the unstable scenarios include $(\lambda{=}0.3, \mu{=}0.6)$ , $(\lambda{=}0.3, \mu{=}0.7)$ , $(\lambda{=}0.4, \mu{=}0.6)$ , $(\lambda{=}0.4, \mu{=}0.7)$ , $(\lambda{=}0.4, \mu{=}0.8)$ , $(\lambda{=}0.4, \mu{=}0.9)$ , $(\lambda{=}0.5, \mu{=}0.6)$ , $(\lambda{=}0.5, \mu{=}0.7)$ , $(\lambda{=}0.5, \mu{=}0.8)$ , $(\lambda{=}0.5, \mu{=}0.9)$ and $(\lambda{=}0.5, \mu{=}1.0)$ . Furthermore, for all the stable parameter scenarios in each case, a significant change trend can be observed as the parameters vary, which indicates that the SDD strategy is sensitive to the change of scenarios.

Similarly, based on the parameter scenarios and time horizon as mentioned above, the CSB strategy is further applied in the dynamic charging guidance problem. As the parameters $\mu$ and $\lambda$ vary, the maximum EV number in each charging station is obtained, as shown in Fig. 11.

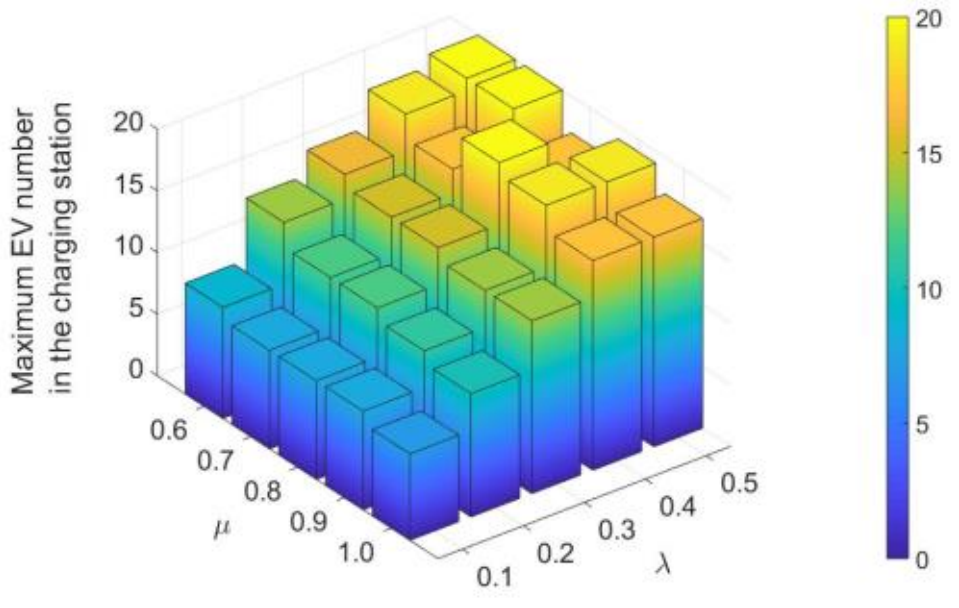

(1) CS 1

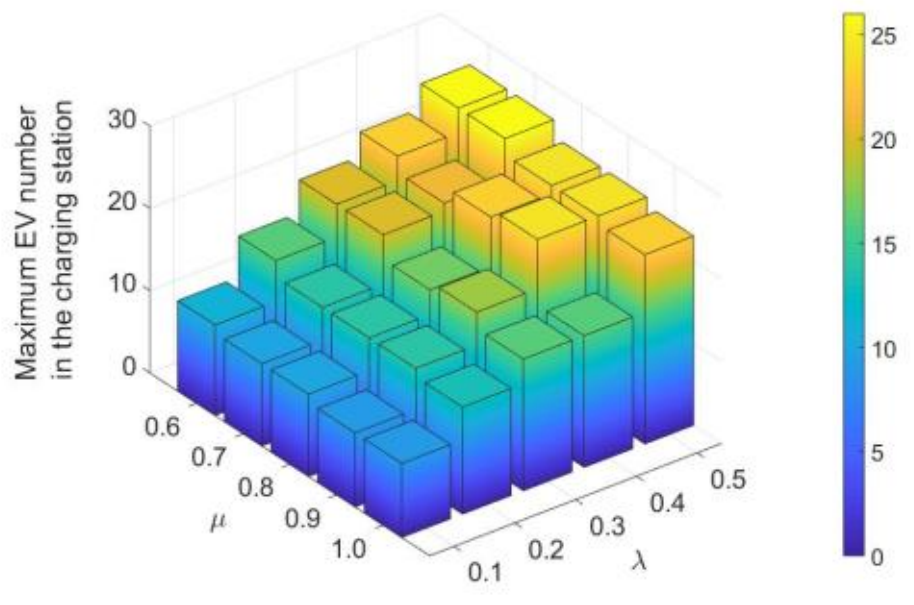

(2) CS 2

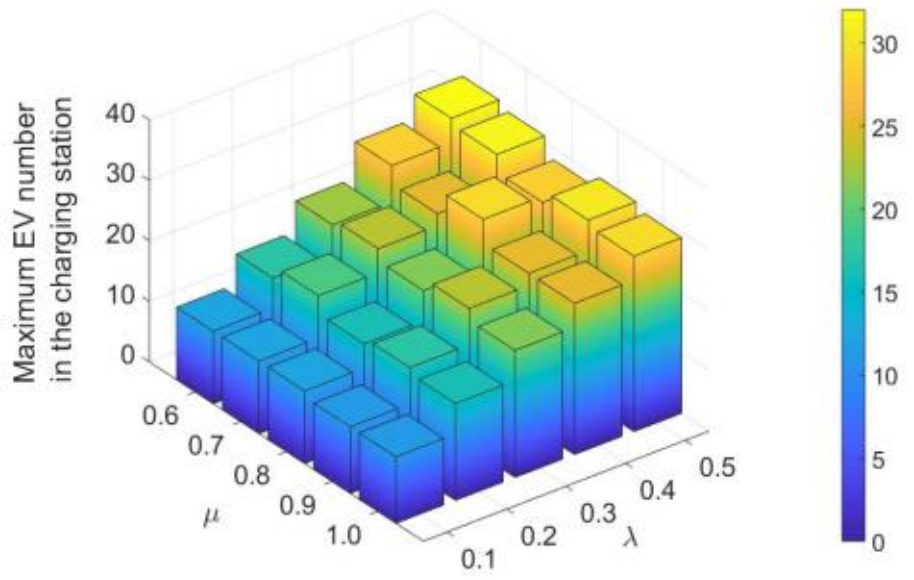

(3) CS 3

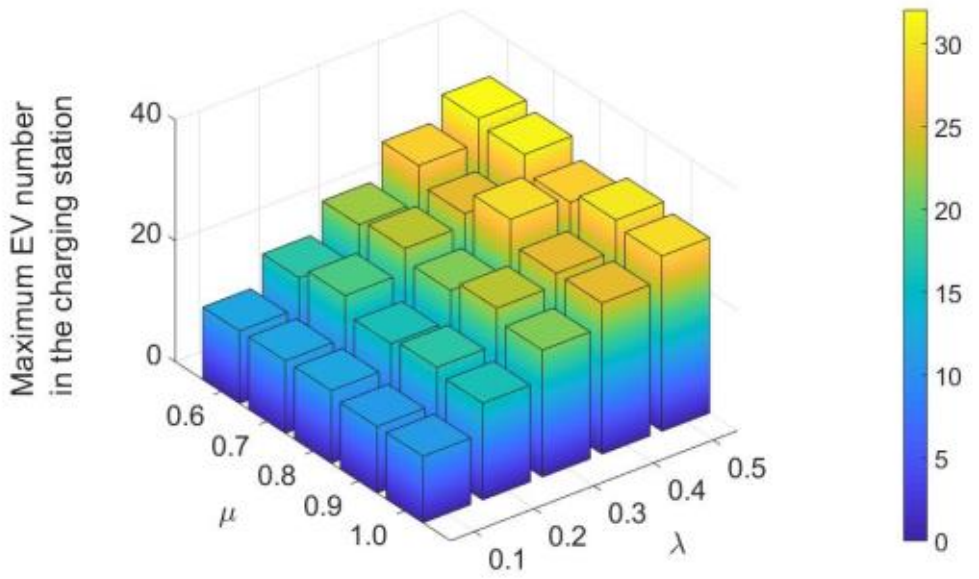

(4) CS 4

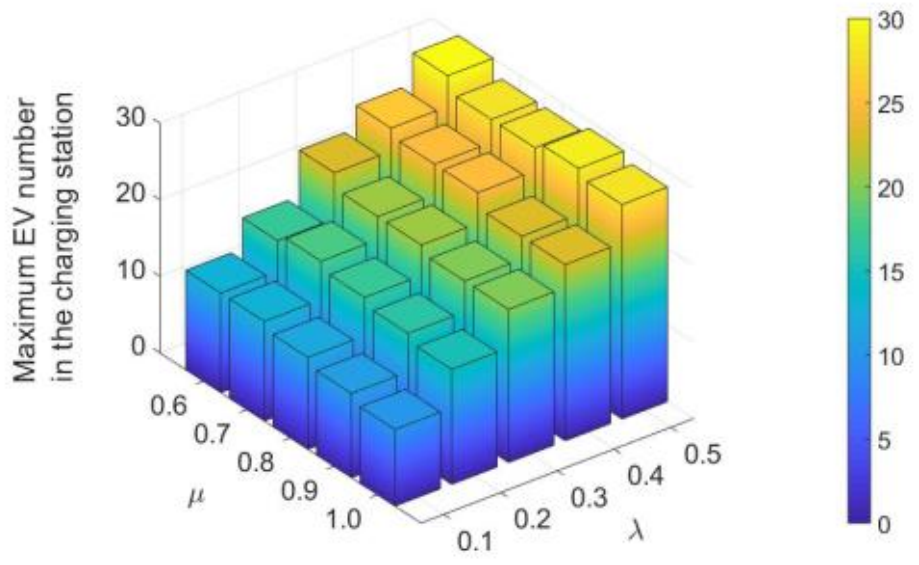

(5) CS 5

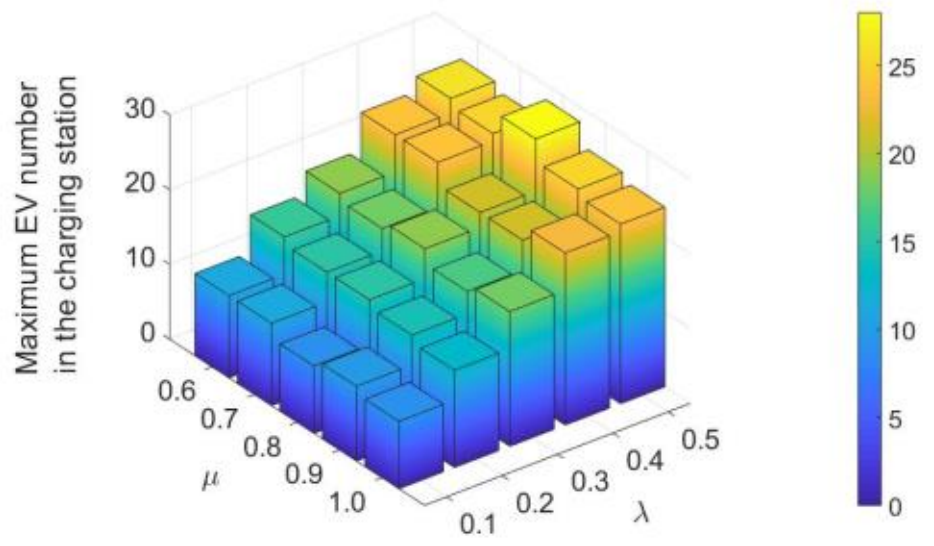

(6) CS 6

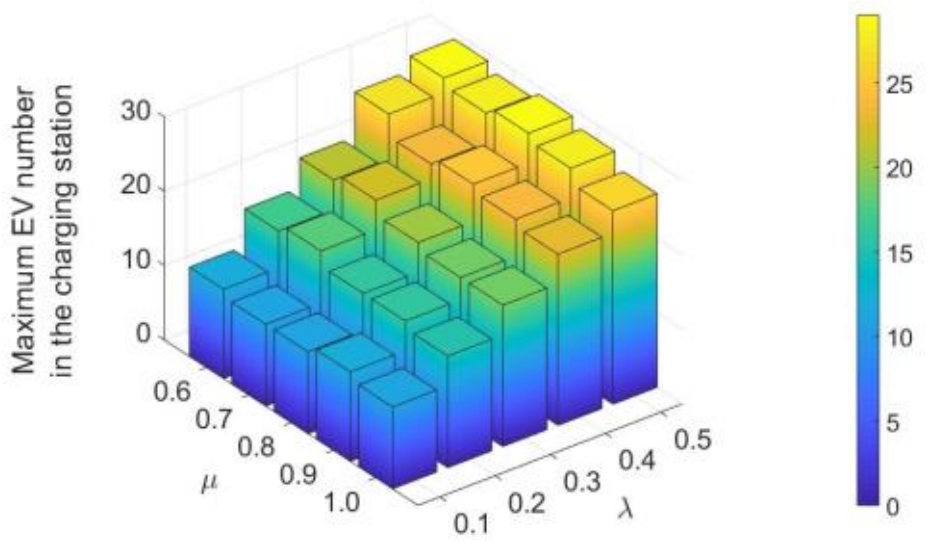

(7) CS 7

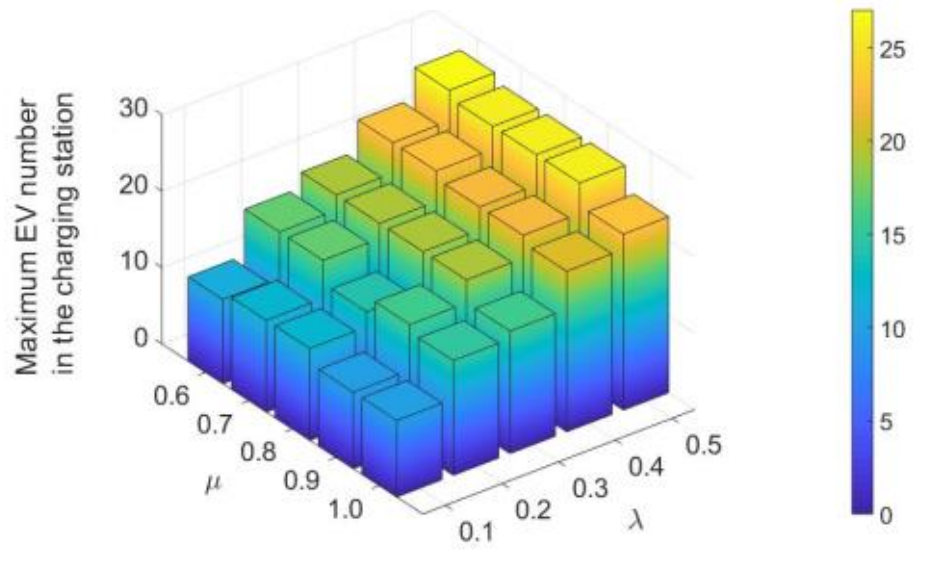

(8) CS 8

Fig.11 Change trends of the maximum EV number in each charging station under the CSB strategy

As can be seen in Fig.11, the case (1)-case (8) respectively present the maximum number of EVs in CS 1-CS 8 under the CSB strategies for the different parameter scenarios. In the case (1), the maximum number of EVs in CS 1 broadly presents a flat increasing trend as the parameter $\mu$ decreases and parameter $\lambda$ increases. However, for some individual scenarios, the moderate fluctuation exists as the parameters vary. The results indicate that the change of scenarios has the relatively limited effects on the CSB strategy as compared to that on the SDD strategy. Among all the parameter scenarios, the peak and lowest values of the maximum EV number in CS 1 respectively equal to 7 and 20. Furthermore, in the case (2)-case (8), the maximum number of EVs in CS 2-CS 8 shows a similar change trend with that in CS 1. Among all the parameter scenarios, the lowest values of the maximum EV number in CS 2-CS 8 are respectively equal to 9, 10, 11, 10, 9, 11 and 10. Comparatively, the peak values of the maximum EV number in CS 2-CS8 equal to 26, 30, 32, 30, 28, 29 and 27, respectively. Moreover, in contrast to the SDD strategy, the CSB strategy has ability to stabilize the state of CS 1-CS8 for all the parameter scenarios, which embodies the advantage of the CSB strategy in terms of the charging station stability.

Through comparing the simulation results in Fig 10 and Fig. 11, it is observed that, for both SDD and CSB strategies, the maximum EV number increases as the parameter $\mu$ decreases and parameter $\lambda$ increases, basically. Given the implication of parameters $\mu$ and $\lambda$ , the simulation results conform to the operation state of charging stations in the real-world situation. To simplify the description, we denote the parameter scenarios with relatively small $\mu$ values and large $\lambda$ values as the “tense charging state”; otherwise, the parameter scenarios are denoted as “relaxed charging state”. As can be seen in the figures, although the SDD and CSB strategies have the similar change trends in terms of the maximum EV number in each charging station, the change degree has the significant difference. When the parameter scenarios vary from “relaxed charging state” to “tense charging state”, under the SDD strategy, a significant change trend of maximum EV number can be observed for each charging station. The maximum number of EVs in each charging station has a significant gap between the adjacent parameter scenarios. As the parameter scenarios further vary, the SDD strategy may have an inability to stabilise the charging service. In contrast, the maximum EV number in each charging station, under the CSB strategy, exhibits a moderate degree of change. For the adjacent parameter scenarios, the maximum EV number has a flat change trend and occasionally presents a moderate fluctuation. Meanwhile, the CSB strategy can stabilise the charging service for all the parameter scenarios. Furthermore, in the “relaxed charging state”, the maximum number of EVs under SDD strategy is close to that under the CSB strategy. However, when the parameter scenarios become the “tense charging state”, a significant gap would exist in the maximum EV number under SDD and CSB strategies. Specifically, for the CSB strategy, the maximum EV number in each charging station is less than 32 and most of the charging stations have the maximum EV number that ranges from 25 to 30. The distribution of the maximum EV number among different charging stations is relatively balanced. On the contrary, for the SDD strategy, the maximum EV number in different charging stations has the significant gap. On the premise of charging station stability, the maximum EV number in different charging stations ranges from 30 to 100, approximately. Therefore, based on the comparing analysis above, it is recommend that, if the charging service has the “tense charging state”, the CSB strategy can be adopted to ensure the operation efficiency of charging stations and service satisfaction of drivers. In contrast, if charging service has the “relaxed charging state”, the SDD strategy would be employed to solve the dynamic charging requests due to its similar effects on charging stations with CSB strategy and consideration of the travel demands of EV drivers.

## 5. Conclusions

To realize the smart charging service for the real-world complicated situation, we propose two optimal strategies to provide guidance for EV charging by considering the dynamic charging requests in a time-varying road network. Based on the analysis in terms of the dynamic characteristic of charging requests, a charging guidance problem is formulated by combining the time-varying road network. Specifically, the dynamic characteristic of charging requests consists in the uncertain aspects regarding time dimension for whether the charging requests occur and their detailed information. The information of charging requests refers to the drivers’ travel destination and remaining energy of EVs. Aiming at the dynamic charging guidance problem, the optimal strategies are established from two different perspectives, including the SDD and CSB strategies. The SDD strategy uses the driving distance from charging stations to destinations as the optimization criterion, based on the travel demands of EV drivers. In contrast, considering the impacts of EV number on the charging station operation and service satisfaction of drivers, the CSB strategy selects the reachable charging stations with minimum vehicle number as the optimal ones, based on the vehicle balance in charging stations. More importantly, despite of the existence of differences between the two strategies, both SDD and CSB strategies have the ability to ensure the reachability of selected charging stations in a time-varying road network. Moreover, the simulation experiments are presented to investigate the performance of the proposed charging guidance strategies. The results of the experiments indicate that, under the identical example scenario, the CSB strategy has a significant advantage to balance the vehicle number among different charging stations as compared to the SDD strategy, especially in the situation with relatively long time horizon. Furthermore, the performance of the two strategies under different scenarios is explored by changing the scenario parameters in the simulation experiments. For the situation with “tense charging state”, the CSB strategy has a better performance than the SDD strategy to ensure the operation efficiency of charging stations and service satisfaction of drivers. Thus, the CSB strategy is suggested to be used in such a situation. On the contrary, the SDD strategy is recommend to be adopted in the situation with “relaxed charging state”, because it has similar effects on the charging station with CSB strategy and further considers the travel demands of EV drivers.

Notably, to simplify the problem formulation, it is assumed that, in every time slot, each normal node can generate at most one charging request. Such an assumption has the certain reasonability if each time slot has the relatively short duration. However, as the scale of EVs increases in the urban transportation system, multiple charging requests may simultaneously occur in the same location of road network, which would significantly complicate the solving processes for the dynamic charging requests. Therefore, built upon the proposed strategies, the distribution rule regarding to the number of charging request occurrence will be further investigated in the future work and considered in extending the charging guidance strategies.

## Acknowledgements

This research is supported by the National Key R&D Program of China (2018YFC0706005, 2018YFC0706000) and National Natural Science Foundation of China (71961137008). The authors also gratefully acknowledge fruitful discussions with Dr. Bin Li in the University of Rhode Island, as well as financial support from the China Scholarship Council.

## References

Abousleiman, R., Rawashdeh, O., 2015. A Bellman-Ford approach to energy efficient routing of electric vehicles. In 2015 IEEE Transportation Electrification Conference and Expo (ITEC), 1-4. IEEE.

Alizadeh, M., Wai, H., Scaglione, A., Goldsmith, A., Fan, Y., Javidi, T., 2014. Optimized path planning for electric vehicle routing and charging. In 2014 52nd Annual Allerton Conference on Communication, Control, and Computing (Allerton), 25-32. IEEE.

Artmeier, A., Haselmayr, J., Leucker, M., Sachenbacher, M., 2010. The shortest path problem revisited: Optimal routing for electric vehicles. In Annual conference on artificial intelligence, 309-316. Springer, Berlin, Heidelberg.

Bi, J., Wang, Y., Sai, Q., Ding, C., 2019. Estimating remaining driving range of battery electric vehicles based on real-world data: A case study of Beijing, China. Energy, 169, 833-843.

Bi, J., Wang, Y., Sun, S., Guan, W., 2018a. Predicting Charging Time of Battery Electric Vehicles Based on Regression and Time-Series Methods: A Case Study of Beijing. Energies, 11(5), 1040.

Bi, J., Wang, Y., Zhang, J., 2018b. A data-based model for driving distance estimation of battery electric logistics vehicles. EURASIP Journal on Wireless Communications and Networking, 2018(1), 251.

Bell, M., Kurauchi, F., Perera, S., Wong, W., 2017. Investigating transport network vulnerability by capacity weighted spectral analysis. Transportation Research Part B: Methodological, 99, 251-266.

Braekers, K., Ramaekers, K., Van Nieuwenhuyse, I., 2016. The vehicle routing problem: State of the art classification and review. Computers & Industrial Engineering, 99, 300-313.

Cao, Y., Tang, S., Li, C., Zhang, P., Tan, Y., Zhang, Z., Li, J., 2011. An optimized EV charging model considering TOU price and SOC curve. IEEE Transactions on Smart Grid, 3(1), 388-393.

Chow, J., Regan, A., 2011. Network-based real option models. Transportation Research Part B: Methodological, 45(4), 682-695.

Daina, N., Sivakumar, A., Polak, J., 2017. Electric vehicle charging choices: Modelling and implications for smart charging services. Transportation Research Part C: Emerging Technologies, 81, 36-56.

De Weerdt, M., Stein, S., Gerding, E., Robu, V., Jennings, N., 2015. Intention-aware routing of electric vehicles. IEEE Transactions on Intelligent Transportation Systems, 17(5), 1472-1482.

Fernandez, L., San Román, T., Cossent, R., Domingo, C., Frias, P., 2010. Assessment of the impact of plug-in electric vehicles on distribution networks. IEEE transactions on power systems, 26(1), 206-213.

Fernández, R., 2018. A more realistic approach to electric vehicle contribution to greenhouse gas emissions in the city. Journal of Cleaner Production, 172, 949-959.

Fiori, C., Ahn, K., Rakha, H., 2018. Optimum routing of battery electric vehicles: Insights using empirical data and microsimulation. Transportation Research Part D: Transport and Environment, 64, 262-272.

Franke, T., Krems, J.F., 2013. Understanding charging behaviour of electric vehicle users. Transportation Research Part F: Traffic Psychology and Behaviour, 21, 75-89.

Fu, L., Sun, D., Rilett, L.R., 2006. Heuristic shortest path algorithms for transportation applications: state of the art. Computers & Operations Research, 33(11), 3324-3343.

Gao, S., Frejinger, E., Ben-Akiva, M., 2010. Adaptive route choices in risky traffic networks: A prospect theory approach. Transportation research part C: emerging technologies, 18(5), 727-740.

Gendreau, M., Ghiani, G., Guerriero, E., 2015. Time-dependent routing problems: A review. Computers & operations research, 64, 189-197.

Gnann, T., Funke, S., Jakobsson, N., Plötz, P., Sprei, F., Bennehag, A., 2018. Fast charging infrastructure for electric vehicles: Today's situation and future needs. Transportation Research Part D: Transport and Environment, 62, 314-329.

He, F., Yin, Y., Lawphongpanich, S., 2014. Network equilibrium models with battery electric vehicles. Transportation Research Part B: Methodological, 67, 306-319.

Huang, Y., Zhao, L., Van Woensel, T. Gross, J., 2017. Time-dependent vehicle routing problem with path flexibility. Transportation Research Part B: Methodological, 95, 169-195.

Huber, G., Bogenberger, K., 2015. Long-Trip Optimization of Charging Strategies for Battery Electric Vehicles. Transportation Research Record, 2497(1), 45-53.

Hung, Y., Michailidis, G., 2012. Stability and control of acyclic stochastic processing networks with shared resources. IEEE Transactions on Automatic Control, 57(2), 489-494.

Hung, Y., Michailidis, G., 2015. Optimal routing for electric vehicle service systems. European Journal of Operational Research, 247(2), 515-524.

International Energy Agency, 2017. Global EV Outlook 2017; IEA: Paris, France.

Jafari, E., Boyles, S., 2017. Multicriteria stochastic shortest path problem for electric vehicles. Networks and Spatial Economics, 17(3), 1043-1070.

Jiang, N., Xie, C., Duthie, J., Waller, S., 2014. A network equilibrium analysis on destination, route and parking choices with mixed gasoline and electric vehicular flows. EURO Journal on Transportation and Logistics, 3(1), 55-92.

Kobayashi, Y., Kiyama, N., Aoshima, H., Kashiyama, M., 2011. A route search method for electric vehicles in consideration of range and locations of charging stations. In 2011 IEEE Intelligent Vehicles Symposium (IV), 920-925. IEEE.

Li, B., Eryilmaz, A., 2014. Non-derivative algorithm design for efficient routing over unreliable stochastic networks. Performance Evaluation, 71, 44-60.

Liu, C., Wu, J., Long, C., 2014. Joint charging and routing optimization for electric vehicle navigation systems. IFAC Proceedings Volumes, 47(3), 2106-2111.

Melliger, M., Van Vliet, O., Liimatainen, H., 2018. Anxiety vs reality–Sufficiency of battery electric vehicle range in Switzerland and Finland. Transportation Research Part D: Transport and Environment. 65, 101-115.

Meng, Q., Yang, H., 2002. Benefit distribution and equity in road network design. Transportation Research Part B: Methodological, 36(1), 19-35.

Neaimeh, M., Hill, G., Hübner, Y., Blythe, P., 2013. Routing systems to extend the driving range of electric vehicles. IET Intelligent Transport Systems, 7(3), 327-336.

Neely, M., Modiano, E., Rohrs, C., 2005. Dynamic power allocation and routing for time-varying wireless networks. IEEE Journal on Selected Areas in Communications, 23(1): 89- 103.

Polson, N., Sokolov, V., 2017. Deep learning for short-term traffic flow prediction. Transportation Research Part C: Emerging Technologies, 79, 1-17.

Putrus, G., Suwanapingkarl, P., Johnston, D., Bentley, E., Narayana, M., 2009. Impact of electric vehicles on power distribution networks. In 2009 IEEE Vehicle Power and Propulsion Conference, 827-831. IEEE.

Qian, K., Zhou, C., Allan, M., Yuan, Y., 2010. Modeling of load demand due to EV battery charging in distribution systems. IEEE Transactions on Power Systems, 26(2), 802-810.

Qin, H., Zhang, W., 2011. Charging scheduling with minimal waiting in a network of electric vehicles and charging stations. In Proceedings of the Eighth ACM international workshop on Vehicular inter-networking, 51-60. ACM.

Raslavičius, L., Azzopardi, B., Keršys, A., Starevičius, M., Bazaras, Ž., Makaras, R., 2015. Electric vehicles challenges and opportunities: Lithuanian review. Renewable and Sustainable Energy Reviews, 42, 786-800.

Rezvani, Z., Jansson, J., Bodin, J., 2015. Advances in consumer electric vehicle adoption research: A review and research agenda. Transportation research part D: transport and environment. 34, 122-136.

Said, D., Cherkaoui, S., Khoukhi, L., 2013. Queuing model for EVs charging at public supply stations. In 2013 9th International Wireless Communications and Mobile Computing Conference (IWCMC), 65-70. IEEE.

Storandt, S., 2012. November. Quick and energy-efficient routes: computing constrained shortest paths for electric vehicles. In Proceedings of the 5th ACM SIGSPATIAL international workshop on computational transportation science, 20-25. ACM.

Strehler, M., Merting, S., Schwan, C., 2017. Energy-efficient shortest routes for electric and hybrid vehicles. Transportation Research Part B: Methodological, 103, 111-135.

Sun, Z., Zhou, X., 2016. To save money or to save time: Intelligent routing design for plug-in hybrid electric vehicle. Transportation Research Part D: Transport and Environment, 43, 238-250.

Sweda, T., Dolinskaya, I., Klabjan, D., 2017. Adaptive routing and recharging policies for electric vehicles. Transportation Science, 51(4), 1326-1348.

Wang, T., Cassandras, C., Pourazarm, S., 2014. Energy-aware vehicle routing in networks with charging nodes. IFAC Proceedings Volumes, 47(3), 9611-9616.

Wang, Y., Bi, J., Guan W., Zhao X., 2018a. Optimising route choices for the travelling and charging of battery electric vehicles by considering multiple objectives. Transportation Research Part D: Transport and Environment. 64, 246-261.

Wang, Y., Bi, J., Zhao, X. Guan, W., 2018b. A geometry-based algorithm to provide guidance for electric vehicle charging. Transportation Research Part D: Transport and Environment, 63, 890-906.

Wang, Y., Jiang, J., Mu, T., 2013. Context-aware and energy-driven route optimization for fully electric vehicles via crowdsourcing. IEEE Transactions on Intelligent Transportation Systems, 14(3), 1331-1345.

Xie, C., Jiang, N., 2016. Relay requirement and traffic assignment of electric vehicles. Computer-Aided Civil and Infrastructure Engineering, 31(8), 580-598.

Xu, M., Meng, Q., Liu, K., 2017. Network user equilibrium problems for the mixed battery electric vehicles and gasoline vehicles subject to battery swapping stations and road grade constraints. Transportation Research Part B: Methodological, 99, 138-166.

Yagcitekin, B., Uzunoglu, M., 2016. A double-layer smart charging strategy of electric vehicles taking routing and charge scheduling into account. Applied energy, 167, 407-419.

Yang, S., Cheng, W., Hsu, Y., Gan, C., Lin, Y., 2013. Charge scheduling of electric vehicles in highways. Mathematical and Computer Modelling, 57(11-12), 2873-2882.

Yi, Z., Bauer, P., 2018. Optimal stochastic eco-routing solutions for electric vehicles. IEEE Transactions on Intelligent Transportation Systems, 19(12), 3807-3817.

Zhang, S., Luo, Y., Li, K., 2016. Multi-objective route search for electric vehicles using ant colony optimization. In 2016 American Control Conference (ACC), 637-642. IEEE.

Zhang, Y., Aliya, B., Zhou, Y., You, I., Zhang, X., Pau, G., Collotta, M., 2018. Shortest feasible paths with partial charging for battery-powered electric vehicles in smart cities. Pervasive and Mobile Computing, 50, 82-93.